# Separations of Matroid Freeness Properties

Arnab Bhattacharyya[*]   Elena Grigorescu[†]   Jakob Nordström[‡]   Ning Xie[§]

November 16, 2018

**Abstract**

Properties of Boolean functions on the hypercube that are invariant with respect to linear transformations of the domain are among some of the most well-studied properties in the context of property testing. In this paper, we study the fundamental class of linear-invariant properties called *matroid freeness* properties. These properties have been conjectured to essentially coincide with all testable linear-invariant properties, and a recent sequence of works has established testability for increasingly larger subclasses of matroid freeness properties. One question that has been left open, however, is whether the infinitely many *syntactically* different matroid freeness properties recently shown to be testable in fact correspond to new, *semantically* distinct properties. This is a crucial issue since it has also been shown previously that there exist subclasses of matroid freeness properties for which an infinite set of syntactically different representations collapse into one of a small, finite set of properties, all previously known to be testable.

An important question is therefore to understand the semantics of matroid freeness properties, and in particular when two syntactically different properties are truly distinct. We shed light on this problem by developing a method for determining the relation between two matroid freeness properties $\mathcal{P}$ and $\mathcal{Q}$. Furthermore, we show that there is a natural subclass of matroid freeness properties such that for any two properties $\mathcal{P}$ and $\mathcal{Q}$ from this subclass, a strong dichotomy must hold: either $\mathcal{P}$ is contained in $\mathcal{Q}$ or the two properties are "well separated" from one another. As an application of this method, we exhibit new, infinite hierarchies of testable matroid freeness properties such that at each level of the hierarchy, there are explicit functions that are far in Hamming distance from all functions lying in the lower levels of the hierarchy. Our key technical tool is an apparently new notion of maps between linear matroids, which we call *labeled matroid homomorphisms*, that might be of independent interest.

## 1 Introduction

The field of property testing, as initiated by [BLR93, BFL91] and defined formally by [RS96, GGR98], asks if, for a given property, there exists an algorithm which queries an input object a small number of times and decides correctly with high probability whether the object has the property or whether it is "far away" from the property. The property is called *testable*, or sometimes *strongly testable* or *locally testable*, if the number of queries can be made independent of the size of the object without affecting the correctness probability. Since such a tester receives only constantly many bits of information about the input object, a prerequisite for testability is that there be quickly detectable local obstructions to the property whenever the input object is far from satisfying it. Perhaps quite surprisingly, it has been found that a large number of different natural properties satisfy this strong requirement and indeed admit property testers (see, for instance, the recent surveys [Ron08, Ron09] for more information). This raises the question what lies behind all of these testability results, and whether we can gain an understanding of some common underlying traits by investigating the space of testable properties.

[*]MIT CSAIL, abhatt@mit.edu. Supported in part by a DOE Computational Science Graduate Fellowship and NSF Awards 0514771, 0728645, and 0732334.

[†]MIT CSAIL, elena_g@csail.mit.edu. Supported in part by NSF award CCR-0829672.

[‡]MIT CSAIL, jakobn@csail.mit.edu. Supported by the Royal Swedish Academy of Sciences, the Ericsson Research Foundation, the Sweden-America Foundation, the Foundation Olle Engkvist Byggmästare, the Sven and Dagmar Salén Foundation, and the Foundation Blanceflor Boncompagni-Ludovisi, née Bildt.

[§]MIT CSAIL, ningxie@csail.mit.edu. Supported by NSF Awards 0514771, 0728645 and 0732334.

## 1.1 Linear Invariance and Matroid Freeness Properties

One particular class of properties of interest is the set of *linear-invariant* properties of functions on the Boolean hypercube. We refer to the recent survey by Sudan [Sud10] for an in-depth discussion of linear invariance and more generally of the relation between invariance and property testing. Briefly, a linear-invariant property $\mathcal{F}$ is a collection of functions $\bigcup_{n=1}^{\infty}\{f : \mathbb{F}^n \to \mathcal{R}\}$, where $\mathbb{F}$ is a finite field and $\mathcal{R}$ is some finite range, such that if $f$ is in $\mathcal{F}$, then $f \circ L$ is also in $\mathcal{F}$ for every $\mathbb{F}$-linear map $L : \mathbb{F}^n \to \mathbb{F}^n$, where $\circ$ denotes function composition $(f \circ L)(x) = f(L(x))$. In this work, we will mostly restrict ourselves to the most commonly studied case $\mathbb{F} = \mathbb{F}_2$ and $\mathcal{R} = \{0, 1\}$.

As first explicitly pointed out by Kaufman and Sudan [KS08], a wide range of natural algebraic properties, whose testability had been previously studied as special cases, are linear-invariant. Examples include such landmark results in the literature as the testability of linear functions [BLR93], Reed-Muller codes [BFL91, RS96, JPRZ04, AKK$^+$05], and BCH codes [KL06]. In view of this, the goal of a sequence of recent papers [BCSX09, KSV10, KSV09, Sha09, BGS10, BCSX10] has been to try to explain these previous results in a uniform way by providing a general understanding of necessary and sufficient conditions for testability of linear-invariant properties. The focus of these works has been on so-called matroid freeness properties,[1] as defined next.

**Definition 1.1 (Matroid freeness).** Given integers $k, r \geq 1$, a set $M = \{\mathbf{v}_1, \ldots, \mathbf{v}_k\}$ of $k$ vectors in $\mathbb{F}^r$, and $\sigma$ a string in $\mathcal{R}^k$, we say that a function $f : \mathbb{F}^n \to \mathcal{R}$ is $(M, \sigma)$-*free* if there does not exist any linear map $L : \mathbb{F}^r \to \mathbb{F}^n$ such that $f(L(\mathbf{v}_i)) = \sigma_i$ for all $i \in [k]$. Otherwise, if such an $L$ exists we say $f$ contains $(M, \sigma)$ at $L$.

Observe that for the purposes of this definition, the exact identity of the elements of $M$ is not important; it is only important to know about the linear dependencies between them. Therefore, it is convenient to consider $M$ abstractly as a linear matroid.[2] With this in view, we refer to the pair $(M, \sigma)$ in the above definition as a *labeled matroid* or a *matroid constraint*.

Definition 1.1 can be generalized so that we have a collection of matroid constraints, instead of just one as above.

**Definition 1.2.** Given a (possibly infinite) collection $\mathbb{M} = \{(M^1, \sigma^1), (M^2, \sigma^2), \ldots\}$ of matroid constraints, a function $f : \mathbb{F}^n \to \mathcal{R}$ is said to be $\mathbb{M}$-free if it is $(M^i, \sigma^i)$-free for all $i$.

The class of $\mathbb{M}$-freeness properties has turned out to be very convenient for a general analysis of the testability of linear-invariant properties. Work along these lines was initiated by [BCSX09] who showed that $(M, \sigma)$-freeness for functions $f : \mathbb{F}_2^n \to \{0, 1\}$ is testable for so-called *graphic matroids* $M$ (as defined in Section 2), provided that $\sigma$ is the all-ones string, henceforth denoted $\sigma = 1^*$. It is easy to verify that such a pattern corresponds to a monotone property, i.e., that if $f$ is $(M, 1^*)$-free and $f'$ is obtained from $f$ by flipping $f(\mathbf{x})$ from 1 to 0 at any point $\mathbf{x} \in \mathbb{F}_2^n$, then $f'$ is also $(M, 1^*)$-free. The papers [KSV10] and [Sha09] independently showed that the restriction to graphic matroids could be dropped, but this again requires that $\sigma = 1^*$. More recently, [BGS10] showed that if $\mathbb{M} = \{(M^1, \sigma^1), (M^2, \sigma^2), \dots\}$ is a possibly infinite collection of matroid constraints such that each $M^i$ is a graphic matroid, but where $\sigma^i$ is no longer necessarily $1^*$, then $\mathbb{M}$-freeness is testable. Furthermore, [BGS10] conjectured that the class of properties captured in Definition 1.2 above is exactly the class of linear-invariant properties testable with one-sided testers. In another work [BCSX10], it was conjectured that properties testable by so-called proximity-oblivious testers, a notion introduced in [GR09], are exactly the class of $\mathbb{M}$-freeness properties with finite $\mathbb{M}$. The important role of the $(M, \sigma)$-freeness properties as the building blocks of potentially all testable linear-invariant properties thus provides a strong motivation to study their structure more carefully.

---
[1]It should be noted that strictly speaking, the properties in [BGS10, KSV10, Sha09] are described in terms of forbidding solutions to systems of linear equations rather then requiring the absence of matroid patterns. These two formulations are essentially equivalent, however, as explained in Appendix A.

[2]The formal definition of a matroid is not too important in this context, so although we provide a definition in Appendix A for completeness, in the rest of this paper the reader can just think of a matroid as a set of elements in a vector space over $\mathbb{F}$.

Even though matroid constraints seem to arise naturally when characterizing the testability of linear-invariant properties, there is potentially a very problematic loophole in claims such as the one above from [BGS10]. The issue is that although the claim shows the testability of $\mathbb{M}$-freeness for infinitely many matroid constraint collections $\mathbb{M}$, it is not at all clear whether these infinitely many $\mathbb{M}$'s characterize infinitely many distinct properties. Specifically, it could well be the case that the property of $(M^1, \sigma^1)$-freeness is identical to $(M^2, \sigma^2)$-freeness for two different matroid constraints $(M^1, \sigma^1)$ and $(M^2, \sigma^2)$. A little more subtly, it could also be the case that for distinct constraints $(M^1, \sigma^1)$ and $(M^2, \sigma^2)$, any function which is $(M^1, \sigma^1)$-free is very close to being $(M^2, \sigma^2)$-free, so that even though the properties are not identical, any one-sided tester for $(M^1, \sigma^1)$-freeness can be easily modified to be a tester for $(M^2, \sigma^2)$-freeness. A third pitfall is that it could be the case that for three distinct matroid constraints $(M^1, \sigma^1)$, $(M^2, \sigma^2)$ and $(M^3, \sigma^3)$, the property of $(M^3, \sigma^3)$-freeness is (or is very close to) the union of the properties $(M^1, \sigma^1)$-freeness and $(M^2, \sigma^2)$-freeness. In this case, testability of $(M^3, \sigma^3)$-freeness is trivially guaranteed by the testability of $(M^1, \sigma^1)$-freeness and $(M^2, \sigma^2)$-freeness. Thus, for the above cited result of [BGS10] to be nontrivial, one should ensure that the properties covered in that result are not the union of properties already previously known to be testable.

It should be stressed that these concerns are far from hypothetical. It was shown in [BCSX09] that if $M$ is the graphic matroid on the $k$-cycle for any $k$, then while there is an infinite hierarchy of distinct properties when $\sigma = 1^k$, it turns out that $(M, \sigma)$-properties when $\sigma \neq 1^k$ always degenerate to one of a finite set of properties that have already been known to be testable since the work of [BLR93]. It is a natural question to ask whether it could be the case more generally that all non-monotone matroid freeness properties degenerate to one of a small set of already well-studied properties. This is posed as an open problem in [BCSX09], and resolving this question is the main motivation behind this work.

## 1.2 Summary of Our Results

Very briefly, given two matroid constraints $(M, \sigma)$ and $(N, \tau)$, we establish necessary and sufficient conditions for when the two properties $(M, \sigma)$-freeness and $(N, \tau)$-freeness are identical or distinct, provided that the constraints satisfy certain structural conditions. We then go on to show the existence of matroid constraints that satisfy these conditions. Finally, we use these results to rule out the aforementioned objections about testability results for matroid freeness properties by exhibiting infinite hierarchies of distinct non-monotone and testable matroid freeness properties. We now describe these results in some more detail.

The main tool we use to show separations between matroid freeness properties is the notion of a *labeled matroid homomorphism*. Just as the notions of graph homomorphisms and its variants are helpful in counting occurrences of (induced) subgraphs inside graphs (see [AS06] for a survey), labeled matroid homomorphisms allow us to count the number of times a given matroid constraint is contained in a function. More precisely, we define a labeled matroid homomorphism $\phi$ from a matroid constraint $(M, \sigma)$ to a matroid constraint $(N, \tau)$ to be a map $\phi$ that *(i)* is linear, *(ii)* maps elements of $M$ to elements of $N$, and *(iii)* preserves labels in the sense that the $\sigma$-label of any element $\mathbf{v}$ in $M$ equals the $\tau$-label of $\mathbf{w} = \phi(\mathbf{v})$ in $N$. Observe that since $\phi$ is linear, if some some elements of $M$ are linearly dependent, then their images in $N$ are also linearly dependent.

It is not hard to show that if there is a labeled matroid homomorphism from $(M, \sigma)$ to $(N, \tau)$, then any $(M, \sigma)$-free function is also $(N, \tau)$-free. It is reasonable to wonder whether the fact that $(M, \sigma)$ does *not* map homomorphically into $(N, \tau)$ can also provide some information about the relationship between the two properties of $(M, \sigma)$-freeness and $(N, \tau)$-freeness. If we are optimistically inclined, we might even ask whether the existence or non-existence of homomorphisms *exactly* determines the relationship between these two properties in the sense that $(M, \sigma)$-freeness is far in Hamming distance from being contained in $(N, \tau)$-freeness in this latter case. Somewhat surprisingly, it turns out that this is in fact true for monotone matroid freeness properties.

**Theorem 1.3 (First main theorem (informal)).** *For any linear labeled matroids $(M, 1^*)$ and $(N, 1^*)$ it holds that either $(M, 1^*)$-freeness is contained in $(N, 1^*)$-freeness or $(M, 1^*)$-freeness is "well separated" from $(N, 1^*)$-freeness in the sense that there is a function that is $(M, 1^*)$-free but far from being $(N, 1^*)$-free. The first case applies if there exists a labeled matroid homomorphism from $(M, 1^*)$ to $(N, 1^*)$, and the second case applies otherwise.*

Notice that this implies a strong dichotomy: one of the two cases in Theorem 1.3 must hold, and it can never be the case that the two properties are distinct but close in a property testing sense.

It should be noted that some limited results of the same flavor were proven for specific graphic matroids and monotone patterns in [BCSX09], and our techniques are inspired by that paper. However, our results are much stronger in that they apply to arbitrary (non-graphic) linear matroids and provide an exact criterion for when two matroid constraints with monotone patterns are distinct.

An obvious next question is whether this dichotomy, and the characterization in terms of labeled matroid homomorphisms, holds not only for monotone properties but for matroid freeness properties in general. Extending our methods to non-monotone properties presents significant technical hurdles, and indeed the general question remains a challenging open problem. However, we are able to identify two special cases when an exact characterization in terms of homomorphisms still applies. We restrict ourselves in the theorem statement below to *graphic* matroid freeness properties, which are of special interest since they were the ones shown to be testable in [BGS10],[3] although our results hold in slightly larger generality and in particular extend even to properties not currently known to be testable.

**Theorem 1.4 (Second main theorem (informal)).** *If $(M, \sigma)$ and $(N, \tau)$ are graphic matroids that satisfy certain structural conditions but where $\sigma$ and $\tau$ can be non-monotone patterns, then it holds that either $(M, \sigma)$-freeness is contained in $(N, \tau)$-freeness or it is "well separated" from it, and this is exactly determined by the existence or non-existence of a labeled matroid homomorphism from $(M, \sigma)$ to $(N, \tau)$.*

Our focus in Theorem 1.4 is on matroids over complete graphs $K_k$, which in a sense are the building blocks of all labeled graphic matroids (again, we refer to Section 2 for a more formal discussion). Our technical contribution lies in using the structure of the complete graph to argue that if $(M, \sigma)$ does not embed homomorphically into $(N, \tau)$, where $M, N$ are graphic submatroids of the complete graph and $\sigma, \tau$ have some specific structure, then it is possible to pack into a function many copies of $(N, \tau)$, i.e., many violations of $(N, \tau)$-freeness, while still keeping the function $(M, \sigma)$-free.

Finally, we apply these two dichotomy results to rule out the concerns discussed above about degeneracy of matroid freeness properties.

**Theorem 1.5 (Third main theorem (informal)).** *There are infinite hierarchies of testable, non-monotone and monotone matroid freeness properties such that for each hierarchy, consecutive properties in the hierarchy are well separated. Furthermore, it is not the case that any one of the properties equals (or is close*

---

[3]In fact, [BGS10] shows testability for a strictly larger class of so-called complexity-1 matroids, but we do not want to get into technicalities here.

*to) the union of some subset of the other properties, and the properties provably cannot simply correspond to e.g. the well-studied class of low-degree polynomials.*

Thus, in particular, Theorem 1.5 allows us to conclude that the properties shown to be testable in [BGS10] do indeed constitute a new, infinitely large class of testable linear-invariant properties.

### 1.3 Organization of This Paper

In Section 2, we provide some necessary background and elaborate more in detail on the motivation behind this work. Section 3 is devoted to establishing the dichotomy theorems. In Section 4, we prove the non-existence of labeled matroid homomorphisms, which can then be combined with the results from Section 3 to establish the infinite hierarchies of testable non-monotone properties in Section 5. We conclude in Section 6 by discussion some of the intriguing questions left open by our work. Some background material, which might be useful although not necessary to understand the rest of the paper, is presented in Appendix A for completeness.

## 2 Preliminaries and Motivation

Let $\mathbb{N} = \{0, 1, \ldots\}$ denote the set of natural numbers and let $\mathbb{N}^+ = \mathbb{N} \setminus \{0\}$. Let $n \geq 1$ be a natural number. We write $[n]$ to denote the set $\{1, 2, \ldots, n\}$. We write $\mathbb{F}$ to denote a (finite) field.

Formally speaking, a *property* $\mathcal{P}$ is a subset of functions from domain(s) $\mathcal{D}_n$ to range(s) $\mathcal{R}_n$, that is $\mathcal{P} = \bigcup_{n \in \mathbb{N}^+} \mathcal{P}_n$ where $\mathcal{P}_n \subseteq \bigcup_{n \in \mathbb{N}^+} \{\mathcal{D}_n \to \mathcal{R}_n\}$, but it is customary to suppress $n$ in this notation. Throughout this paper, we will have $\mathcal{D}_n = \mathbb{F}^n$ and $\mathcal{R}_n = \mathcal{R}$ for some fixed $\mathcal{R}$, usually $\mathcal{R} = \{0, 1\}$.

Let $f, g : \mathcal{D} \to \mathcal{R}$ be two functions defined over the same domain $\mathcal{D}$. The *(relative) distance* between functions $f$ and $g$, denoted $\text{dist}(f, g)$, is the probability $\Pr_{x \in \mathcal{D}}[f(x) \neq g(x)]$ that they differ on some $x$ drawn uniformly at random from $\mathcal{D}$. The distance between a function $f$ and a property $\mathcal{P}$ is $\text{dist}(f, \mathcal{P}) = \min_{g \in \mathcal{P}}\{\text{dist}(f, g)\}$. We say that $f$ is $\delta$-*far* from $\mathcal{P}$ if $\text{dist}(f, \mathcal{P}) \geq \delta$ and $\delta$-*close* otherwise. The following two definitions capture the notion of two properties being "well separated" from each other in a property testing sense.

**Definition 2.1 ($\delta$-separated).** For two properties $\mathcal{P}, \mathcal{Q} \subseteq \bigcup_{n \in \mathbb{N}^+}\{\mathbb{F}^n \to \mathcal{R}\}$, we say that $\mathcal{Q}$ is $\delta$-*separated* from $\mathcal{P}$ if for infinitely many $n$ there are functions $f_n : \mathbb{F}^n \to \mathcal{R}$ that are in $\mathcal{Q}$ but are $\delta$-far from being in $\mathcal{P}$ (where we note that $\delta > 0$ is fixed and in particular independent of $n$).

**Definition 2.2 ($\delta$-strictly contained).** For two properties $\mathcal{P}, \mathcal{Q} \subseteq \bigcup_{n \in \mathbb{N}^+}\{\mathbb{F}^n \to \mathcal{R}\}$, we say that $\mathcal{P}$ is $\delta$-*strictly contained* in $\mathcal{Q}$ if $\mathcal{P} \subseteq \mathcal{Q}$ but $\mathcal{Q}$ is $\delta$-separated from $\mathcal{P}$.

In this work, we will not be too concerned with actual property testing, focusing instead on understanding the semantics of (syntactic) properties already shown to be testable by other means. In order for the discussion in this paper to be self-contained, however, we recall that a *tester* for a property $\mathcal{P}$ is a probabilistic algorithm which is given a distance parameter $\delta$ and has oracle access to an input function $f : \mathcal{D} \to \mathcal{R}$. The tester should accept with high probability, say at least $2/3$, if $f \in \mathcal{P}$ and should reject, also with probability at least $2/3$, if the function is $\delta$-*far* from $\mathcal{P}$. The tester is said to be *one-sided* if it has no false negatives, i.e., if functions in the property are always accepted with probability 1. The central parameter associated with a tester is the number of oracle queries it makes to the function $f$ being tested. In particular, a property is called *testable* (or *locally testable*) if there is a tester with query complexity that depends only on the distance parameter $\delta$ and is independent of the size of the domain $\mathcal{D}$.

Recall that a property $\mathcal{P} \subseteq \bigcup_{n \in \mathbb{N}^+}\{\mathbb{F}^n \to \mathcal{R}\}$ is said to be *linear-invariant* if $f \in \mathcal{P}$ implies that $f \circ L \in \mathcal{P}$ for every linear transformation $L : \mathbb{F}^n \to \mathbb{F}^n$. This notion should not be confused with the well-studied property of being *linear*, i.e., that if it holds for $f, g : \mathbb{F}^n \to \mathbb{F}$ that if $f \in \mathcal{P}$ and $g \in \mathcal{P}$, then it must also be the case that $f + g \in \mathcal{P}$. Note that a linear-invariant property need not be linear. Indeed, in general this will not be the case, and in our setting we will not impose any algebraic structure on the range $\mathcal{R}$ of the functions.

Turning next to matroids, for the purposes of this paper the reader can think of a *linear matroid* $M$ as a set of vectors $\{\mathbf{v}_1, \ldots, \mathbf{v}_k\}$ in $\mathbb{F}^{k'}$ for $k' \leq k$. We will often write $N$ to denote some other matroid $\{\mathbf{w}_1, \ldots, \mathbf{w}_m\}$ over $\mathbb{F}^{m'}$ for $m' \leq m$, and unless otherwise stated $\mathbf{v}_i$ is assumed to denote a vector in $M$ and $\mathbf{w}_j$ to denote a vector in $N$. We let $\mathbf{e}_1, \mathbf{e}_2, \ldots$ denote the unit vectors in the ambient space, i.e., $\mathbf{e}_i$ is 1 at coordinate $i$ and 0 everywhere else. Sometimes when we need to distinguish basis vectors of different matroids we will also write $\mathbf{f}_1, \mathbf{f}_2, \ldots$ to denote unit vectors (in some other space). The weight $|\mathbf{v}|$ of a vector $\mathbf{v}$ is the number of non-zero coordinates in $\mathbf{v}$. We write $\sigma = \langle \sigma_1, \ldots, \sigma_k \rangle \in \mathcal{R}^k$ and $\tau = \langle \tau_1, \ldots, \tau_m \rangle \in \mathcal{R}^m$ to denote strings or patterns corresponding to the matroids $M$ and $N$ respectively. We let $f$ and $g$ denote functions $\mathbb{F}^n \to \mathcal{R}$, where we will often, but not always, have $\mathbb{F} = \mathbb{F}_2 = \mathrm{GF}(2)$ and $\mathcal{R} = \{0, 1\}$. Note that $k, k', m, m'$ are all fixed while we think of $n$ as going to infinity.

A matroid $M = \{\mathbf{v}_1, \ldots, \mathbf{v}_k\}$ is said to be *graphic* if there exists a graph $G$ with $k$ edges for which these edges can be associated with the vectors $\mathbf{v}_1, \ldots, \mathbf{v}_k$ in $M$ in such away that any subset of vectors $S \subseteq \{\mathbf{v}_1, \ldots, \mathbf{v}_k\}$ is linearly dependent if and only if the associated set of edges contains a cycle. In this case, we denote $M$ by $M(G)$. (Also notice that we write $\mathbf{v}$ and $\mathbf{e}$ for vectors to distinguish them from vertices $v$ and edges $e$ in graphs.) We require the graph $G$ to be simple; that is, $G$ has no self-loops or parallel edges. It is a well-known fact that graphic matroids always can be represented as *binary matroids*, i.e., linear matroids over $\mathbb{F}_2$.

A central notion of this work will be that of a *matroid homomorphism*.

**Definition 2.3 (Matroid homomorphism [BCSX09]).** Let $M = \{\mathbf{v}_1, \ldots, \mathbf{v}_k\}$ and $N = \{\mathbf{w}_1, \ldots, \mathbf{w}_m\}$ be two matroids with $M \subseteq \mathbb{F}^{k'}$ and $N \subseteq \mathbb{F}^{m'}$. A *matroid homomorphism* $\phi : M \to N$ is a $\mathbb{F}$-linear map from $\mathbb{F}^{k'}$ to $\mathbb{F}^{m'}$ such that $\phi(\mathbf{v}_i) \in \{\mathbf{w}_1, \ldots, \mathbf{w}_m\}$ for every $1 \leq i \leq k$. We will also say that $\phi$ is an *embedding* of the matroid $M$ into the matroid $N$, or that $M$ *embeds* into $N$.

In contrast to [BCSX09], we want to be able to study not only monotone but also non-monotone matroid freeness properties, i.e., properties characterized by matroid constraints $(M, \sigma)$ where we can have $\sigma \notin \{0^k, 1^k\}$. In order to do so, we need the following generalization of Definition 2.3 to arbitrary matroid constraints.

**Definition 2.4 (Labeled matroid homomorphism).** A *labeled matroid homomorphism* $\phi : (M, \sigma) \to (N, \tau)$ is a matroid homomorphism from $M$ to $N$ which in addition preserves labels in the sense that if $\phi(\mathbf{v}_i) = \mathbf{w}_j$, then $\sigma_i = \tau_j$. If there exists a labeled homomorphism from $(M, \sigma)$ to $(N, \tau)$, we say that $(M, \sigma)$ *embeds* into $(N, \tau)$ and write $(M, \sigma) \hookrightarrow (N, \tau)$; otherwise, we write $(M, \sigma) \not\hookrightarrow (N, \tau)$.

Let us now fix $\mathcal{R} = \{0, 1\}$ for the rest of this section. We can visualize a graphic matroid constraint $(M(G), \sigma)$ as the graph $G$ with 0/1-labels $\sigma_i$ on its edges. In what follows, we will sometimes identify $(M(G), \sigma)$ with this *labeled graph* $(G, \sigma)$. We say that $(G, \sigma)$ is a *labeled subgraph* of $(H, \tau)$ if $G$ is a subgraph of $H$ such that the edge labels of the common edges coincide. For graphic matroid constraints $(M(G), \sigma)$ and $(M(H), \tau)$, a labeled matroid homomorphism is simply a mapping of edges in $G$ to edges with the same labels in $H$ such that cycles in $G$ map to cycles in $H$. Clearly, if $(G, \sigma)$ is a labeled subgraph $(H, \tau)$, then the embedding of the vertices of $G$ in $H$ induces a matroid homomorphism in the natural way. These are not the only labeled graphic homomorphisms, however. In particular, a matroid homomorphism need not map edges incident to the same vertex in $G$ to incident edges in $H$.

As long as we are only studying monotone patterns, we need not worry too much about exactly how our linear matroids are represented. When we want to discuss $(M, \sigma)$-freeness for a non-monotone pattern $\sigma$, however, we have to specify exactly how $M$ is represented in order to make sure that the matroid constraint is well-defined.

**Definition 2.5 (Standard binary matroid representation).** We say that a binary matroid $M = \{\mathbf{v}_1, \ldots, \mathbf{v}_k\}$ is in *standard representation* if there is a $d < k$ such that $\mathbf{v}_i = \mathbf{e}_i$ for $i = 1, \ldots, d$ and the rest of the vectors are linear combinations of these vectors, i.e., for all $j > d$ we have $\mathbf{v}_j = \sum_{i \in I_j} \mathbf{e}_i$ for index sets $I_j \subseteq [d]$, where $\mathbf{v}_j$, $j > d$, are enumerated in lexicographical order with respect to $I_j$.

It is easy to see that any matroid has a standard linear representation, namely by associating unit vectors to a basis of the matroid and then representing the other elements by appropriate linear combinations of the basis elements. When we talk about matroid constraints from now on, it will always be for linear matroids in standard representation.

We remark that the standard representation does not uniquely specify matroid constraints $(M, \sigma)$ — for a fixed $M$ in standard representation there can be several distinct patterns $\sigma^1, \sigma^2, \ldots$ such that $(M, \sigma^i)$ all represent the same labeled matroid. For instance, for the complete graph $K_d$ on $d$ vertices, all labeled matroids $(M(K_d), 1^{d_1} 0 1^{d_2})$ are easily verified to be the same for all $d_1 + d_2 = \binom{d}{2} - 1$. However, it is not the case that for all $d_1 + d_2 + d_3 = \binom{d}{2} - 2$ the labeled matroids $(M(K_d), 1^{d_1} 0 1^{d_2} 0 1^{d_3})$ are all the same. On the contrary, one can prove (although we do not do so here) that one gets two different cases depending on whether the two 0-labeled edges are incident to a common vertex or not. The point is that the standard representation of $M$ gives us at least *one* well-defined description of the labeled matroid, whereas the example just discussed shows that without such a representation a non-monotone labeled matroid $(M, \sigma)$ could be ambiguous.

In this work, we will focus on a particular class of matroid freeness properties, the understanding of which is arguably fundamental to the broader study of the space of testable linear-invariant properties. Let us explain what we mean by this.

It was shown in [BGS10] that any linear-invariant property in $\mathbb{F}_2^n \to \{0, 1\}$ that is testable with a one-sided tester can be written as an $\mathbb{M}$-freeness property for a possible infinite collection $\mathbb{M}$ of matroid constraints.[4] In other words, any such property can be characterized as the intersection of a possibly infinite number of $(M, \sigma)$-freeness properties. Moreover, each $(M, \sigma)$-freeness property can be written as the intersection of a finite collection of $(F_d, \sigma')$-freeness properties, where $F_d$ is the *full linear matroid*, i.e., the full linear space $F_d = \{\sum_{i \in I} \mathbf{e}_i : I \subseteq [d], I \neq \emptyset\}$ of some dimension $d$. Namely, suppose that the matroid $M$ lives in $\mathbb{F}_2^d$ and let us for simplicity assume that it consists of the first $k \leq 2^d - 1$ vectors in the standard representation. Then it is straightforward to verify that $(M, \sigma)$-freeness is exactly the intersection of all $(F_d, \sigma\tau)$-freeness properties, where $\tau$ ranges over all patterns in $\{0, 1\}^{2^d - (k+1)}$ and $\sigma\tau$ denotes concatenation. (This corresponds to that a violation of $(M, \sigma)$-freeness occurs as soon as the first $k$ vectors in $F_d$ are mapped to points in $\mathbb{F}^n$ evaluating to the pattern $\sigma \in \{0, 1\}^k$, regardless of what the evaluation pattern looks like for the rest of the points.) As another example of the expressive power of matroid constraints, note low-degree polynomials (with constant term zero), can be specified as the intersection of all $(F_d, \tau)$-freeness properties, where $d$ is fixed and $\tau$ ranges over all patterns in $\{0, 1\}^{2^d - 1}$ of odd parity.

It follows from the preceding paragraph that all linear-invariant properties in $\mathbb{F}_2^n \to \{0, 1\}$ that are one-sided-testable are collections of full linear matroid freeness properties, so matroid constraints $(F_d, \sigma)$ can be seen to be the building blocks of all one-sided-testable linear-invariant properties. In the other direction, [BGS10] established that any collection $\mathbb{M}$ of *graphic* matroid freeness properties is testable by a one-sided tester. It is again not hard to see that for any graphic matroid contraint $(M(G), \sigma)$, where $G$ is a graph on $d$ vertices, one can represent $(M(G), \sigma)$-freeness as the intersection of all $(M(K_d), \sigma\tau)$-freeness properties where $\sigma$ labels the edges of $G$ and $\tau$ ranges over all possible labels on the edges in $K_d$ not present in $G$.

---

[4]Modulo a standard technical assumption that is not relevant to this discussion — we refer to [BGS10] for the precise statement.

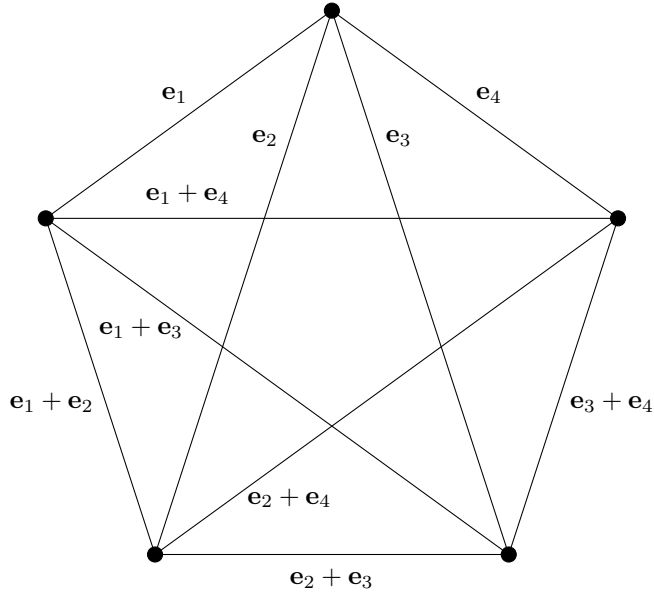

**Figure 1:** The standard matroid representation for $M(K_5)$.

Thus, to gain an understanding of the semantics of matroid freeness properties in general, and of the testability results in [BGS10] in particular, a necessary first step is to comprehend graphic matroid constraints $(M(K_d), \sigma)$. Furthermore, even this first step appears to be a challenging problem in its own right. Therefore, in what follows we will mostly focus on matroids over complete graphs. For the rest of this paper, we fix the representation of such matroids as follows.

**Definition 2.6 (Standard representation of complete graph matroids).** We choose the $d-1$ independent basis vectors of $M(K_d)$ to be the $d-1$ edges incident to some (arbitrarily chosen but) fixed vertex. The $\binom{d}{2}$ vectors in $M(K_d)$ will then consist of all the $d-1$ weight-1 vectors and all the $\binom{d-1}{2}$ weight-2 vectors in $\mathbb{F}_2^{d-1}$. Moreover, these $\binom{d}{2}$ vectors are always ordered lexicographically as

$$M(K_d) = \{\mathbf{e}_1, \mathbf{e}_2, \ldots, \mathbf{e}_{d-1}, \mathbf{e}_1 + \mathbf{e}_2, \mathbf{e}_1 + \mathbf{e}_3, \ldots, \mathbf{e}_1 + \mathbf{e}_{d-1}, \mathbf{e}_2 + \mathbf{e}_3, \mathbf{e}_2 + \mathbf{e}_4, \ldots, \mathbf{e}_{d-2} + \mathbf{e}_{d-1}\} \ .$$

See Figure 1 for an illustration of the standard matroid representation of $M(K_5)$. We make the observation, which will be used later, that for any fixed vector $\mathbf{v} \in M(K_d)$ we can find a standard matroid basis that that contains $\mathbf{v}$ and is on the form in Definition 2.6, i.e., where every non-basis matroid vectors is a sum of two basis vectors. For the weight-1 vectors this is by definition, but it also holds for weight-2 vectors by symmetry. Namely, if we fix a vector $\mathbf{e}_i + \mathbf{e}_j$ for $i < j$, it is immediate to verify that the set of vectors $\{\mathbf{e}_i, \mathbf{e}_i + \mathbf{e}_1, \ldots \mathbf{e}_i + \mathbf{e}_{i-1}, \mathbf{e}_i + \mathbf{e}_{i+1} \ldots \mathbf{e}_i + \mathbf{e}_j, \ldots \mathbf{e}_i + \mathbf{e}_{d-1}\}$ and their pairwise sums generate $M(K_d)$. Another easy way of seeing this is perhaps to take a look at Figure 1 and make a "proof by picture."

As a final notational convention, we make explicit our use of wildcards in patterns, allowing the use of * when the meaning is clear from context. For instance, we will write $1^*$ to denote the all-ones pattern, and $0^d 1^*$ denotes the pattern with ones everywhere except in the first $d$ positions (relative to some fixed representation of the matroid in question).

## 3 A Method for Proving Distinctness of Matroid Freeness Properties

In this section, we develop a method to determine the relations between matroid freeness properties by way of labeled matroid homomorphisms. In brief, we establish a connection between the existence of an embedding from $(M, \sigma)$ to $(N, \tau)$ on the one hand and the distance between the property of being $(M, \sigma)$-free and the property of being $(N, \tau)$-free on the other. Namely, we show that if there exists a labeled homomorphism between $(M, \sigma)$ and $(N, \tau)$ then $(N, \tau)$-freeness contains $(M, \sigma)$-freeness, and that otherwise, at least in some specific cases which we characterize, these properties are well separated in the sense of Definition 2.1. We find this quite surprising, since it is not at all clear a priori how to amplify non-existence of embeddings into statistical distance between the associated properties. We note that it remains an intriguing open problem whether this connection holds in general for any pair of binary matroids $(M, \sigma)$ and $(N, \tau)$.

Let us start with the lemma establishing the easy direction of the connection between labeled matroid homomorphisms and containment of matroid freeness properties.

**Lemma 3.1.** *If $M$ and $N$ are any linear matroids such that there is a labeled matroid homomorphism from $(M, \sigma)$ to $(N, \tau)$, then $(M, \sigma)$-freeness is contained in $(N, \tau)$-freeness.*

Before proving the lemma, we state a simple corollary that will be useful later on.

**Corollary 3.2.** *If $(G, \sigma)$ is a labeled subgraph of $(H, \tau)$ and $f : \mathbb{F}_2^n \to \mathcal{R}$ is a $(M(G), \sigma)$-free function, then $f$ is also $(M(H), \tau)$-free.*

*Proof of Corollary 3.2.* If $(G, \sigma)$ is a labeled subgraph of $(H, \tau)$, then in particular there is a labeled matroid homomorphism from $(M(G), \sigma)$ to $(M(H), \tau)$, namely the one induced by the embedding of the vertices of $G$ in $H$. The claim now follows by Lemma 3.1. □

*Proof of Lemma 3.1.* Let $\phi : M \to N$ be a labeled matroid homomorphism. Suppose that $f : \mathbb{F}^n \to \mathcal{R}$ is not $(N, \tau)$-free and in particular that $f$ contains $(N, \tau)$ at a linear map $L : N \to \mathbb{F}^n$. We claim that this implies that $f$ contains $(M, \sigma)$ at the linear map $L \circ \phi : M \to \mathbb{F}^n$. By assumption, for all $j \in [m]$ we have $f(L(\mathbf{w}_j)) = \tau_j$. Suppose $\phi(\mathbf{v}_i) = \mathbf{w}_{j_i}$. Then by definition $\sigma_i = \tau_{j_i}$ since $\phi$ preserves labels, and for all $i \in [k]$ it clearly holds that $f((L \circ \phi)(\mathbf{v}_i)) = f(L(\phi(\mathbf{v}_i))) = f(L(\mathbf{w}_{j_i}))) = \tau_{j_i} = \sigma_i$, establishing the claim. □

Lemma 3.1 provides a method of arguing that some syntactically different properties are in fact identical. Consider for example $(M(K_3), 1^*)$-freeness and $(M(K_4), 1^*)$-freeness. Since $(K_3, 1^*)$ is a labeled subgraph of $(K_4, 1^*)$, clearly $(M(K_3), 1^*)$-freeness is contained in $(M(K_4), 1^*)$-freeness. Perhaps somewhat counter-intuitively, we can also show that there is a labeled matroid homomorphism from $(M(K_4), 1^*)$ to $(M(K_3), 1^*)$ and hence the containment holds in the other direction as well. To see this, write $M(K_3)$ in standard representation over $\mathbf{e}_1, \mathbf{e}_2$ and $M(K_4)$ in standard representation over $\mathbf{f}_1, \mathbf{f}_2, \mathbf{f}_3$. Define $\phi : M(K_4) \to M(K_3)$ by $\phi(\mathbf{f}_1) = \mathbf{e}_1$, $\phi(\mathbf{f}_2) = \mathbf{e}_2$, and $\phi(\mathbf{f}_3) = \mathbf{e}_1 + \mathbf{e}_2$ and extend it to all of $M(K_4)$ by linearity. We leave it to the reader to verify that $\phi(\mathbf{f}_i + \mathbf{f}_j) \in M(K_3)$ for all $1 \leq i < j \leq 3$. Since $\phi$ is trivially label preserving when all vectors are $1$-labeled, it follows that $\phi$ is a labeled matroid homomorphism. We write this down as a proposition for reference.

**Proposition 3.3.** *The labeled matroid $(M(K_4), 1^*)$ embeds into $(M(K_3), 1^*)$, so $(M(K_3), 1^*)$-freeness and $(M(K_4), 1^*)$-freeness is the same property.*

To provide some more intuition, we give another hopefully instructive example, this time of non-identical properties. Observe that any function which is $(M(K_3), 011)$-free is also $(M(K_4), 011111)$-free by Corollary 3.2, since $(K_3, 011)$ is clearly a labeled subgraph of $(K_4, 011111)$. Also, it is not too hard to show that $(M(K_3), 011)$-freeness and $(M(K_4), 011111)$-freeness are not *exactly* the same. To see this, fix any $\mathbf{y} \in \mathbb{F}_2^n \setminus \{\mathbf{0}\}$ and consider the function $f_{\mathbf{y}} : \mathbb{F}_2^n \to \{0, 1\}$ defined by $f_{\mathbf{y}}(\mathbf{x}) = 1$ if $\mathbf{x} = \mathbf{y}$ and $f_{\mathbf{y}}(\mathbf{x}) = 0$ otherwise. We want to argue that $f_{\mathbf{y}}$ is $(M(K_4), 011111)$-free but not $(M(K_3), 011)$-free. Let again $M(K_3)$ be represented over unit vectors $\mathbf{e}_1, \mathbf{e}_2$ and $M(K_4)$ over unit vectors $\mathbf{f}_1, \mathbf{f}_2, \mathbf{f}_3$. The linear map $L_1 : M(K_3) \to \mathbb{F}_2^n$ sending $\mathbf{e}_1$ to $\mathbf{0}$ and $\mathbf{e}_2$ to $\mathbf{y}$ gives a pattern $\langle 011 \rangle$ for $M(K_3)$. Now suppose that $f_\mathbf{y}$ would contain $(M(K_4), 011111)$ at some linear map $L_2 : M(K_4) \to \mathbb{F}_2^n$. Then in particular $f_\mathbf{y}(L_2(\mathbf{f}_2)) = f_\mathbf{y}(L_2(\mathbf{f}_3)) = 1$, so we must have $L_2(\mathbf{f}_2) = L_2(\mathbf{f}_3) = \mathbf{y}$. But then $f_\mathbf{y}(L_2(\mathbf{f}_2 + \mathbf{f}_3)) = f_\mathbf{y}(L_2(\mathbf{f}_2) + L_2(\mathbf{f}_3)) = f_\mathbf{y}(\mathbf{0}) = 0 \neq 1$. Contradiction. Hence $f_\mathbf{y}$ is $(M(K_4), 011111)$-free.

Note that the above argument shows that $(M(K_3), 011)$-freeness and $(M(K_4), 011111)$-freeness are distinct, but does not rule out that the two properties are "essentially" the same in that they are very close in Hamming distance. However, it follows as a corollary of results that we will prove later in this paper (see Lemma 5.2) that not only does $(M(K_4), 011111)$-freeness contain $(M(K_3), 011)$-freeness, but this containment is strict in the sense of Definition 2.2.

We next consider the other, harder, direction in the correspondence between labeled matroid homomorphisms and matroid freeness properties.

### 3.1 The High-Level Idea

We want to reduce the problem of separating $(M, \sigma)$-freeness from $(N, \tau)$-freeness to a question about matroid homomorphisms. Our goal is to prove a statement of the following kind:

> *Suppose that $(M, \sigma)$ and $(N, \tau)$ are matroid constraints such that there is no labeled matroid homomorphism from $(M, \sigma)$ to $(N, \tau)$. Then the property of $(M, \sigma)$-freeness is $\delta$-separated from that of $(N, \tau)$-freeness.*

To prove such a statement, we need to exhibit an infinite family of functions $f_n : \mathbb{F}^n \to \mathcal{R}$ for $n \to \infty$ that is $(M, \sigma)$-free but far from $(N, \tau)$-free. The general outline of the argument is as follows:

1. First, we define a "canonical function" $f_{N,\tau} : \mathbb{F}^n \to \mathcal{R}$ that encodes the structure of the labeled matroid $(N, \tau)$. More precisely, suppose that $N = \{\mathbf{w}_1, \ldots, \mathbf{w}_m\} \subseteq \mathbb{F}^d$. Then $f_{N,\tau}$ is constructed by splitting $\mathbf{x} \in \mathbb{F}^n$ into $\{\mathbf{y}|\mathbf{z}\}$ for $\mathbf{y} \in \mathbb{F}^d$ and $\mathbf{z} \in \mathbb{F}^{n-d}$ and letting $f_{N,\tau}(\mathbf{x})$ be (in some sense) the indicator function for whether $\mathbf{y} \in \mathbb{F}^d$ is a vector in $N$ and if so what label it has.

2. Then, we prove that $f_{N,\tau}$ is dense in instances of the matroid pattern $(N, \tau)$ and has to be changed in many positions to become $(N, \tau)$-free.

3. Finally, we assume that $f_{N,\tau}$ contains an instance of the matroid $(M, \sigma)$ as witnessed by the linear transformation $L : M \to \mathbb{F}^n$. Then we want to argue that composing $L$ with the projection $\pi$ that maps $\mathbf{x} = \{\mathbf{y}|\mathbf{z}\}$ to $\mathbf{y}$, we obtain a labeled matroid homomorphism $\pi \circ L$ from $(M, \sigma)$ to $(N, \tau)$. But this contradicts the assumption that there is no such homomorphism.

To construct a function that is dense in a pattern is relatively straightforward, and we achieve this (essentially) by a padding argument. The hard part is the third and final step in the argument. Notice that $\pi \circ L$ is linear by construction, but we are only guaranteed that it maps $M$ to $\mathbb{F}^d$, not into $N$. In general, most vectors in $\mathbb{F}^d$ are *not* in $N$, so we somehow have to make sure that we land only in this subset of vectors. Furthermore, if it holds that $(\pi \circ L)(\mathbf{v}_i) = \mathbf{w}_j$, then we must make sure that the labels $\sigma_i$ and $\tau_j$ agree. We remark that in fact, it is not at all clear what it actually means that $f_{N,\tau}$ should be an "indicator function" for $(N, \tau)$. The function $f_{N,\tau}$ has to map all of $\mathbb{F}^n$ to $\mathcal{R}$, and in general for each value $\tau_j \in \mathcal{R}$ there will be some vector $\mathbf{w}_j \in N$ such that $\tau_j$ is the correct value. However, all the (majority of) vectors $\mathbf{x} \in \mathbb{F}^n$ that do *not* correspond to vectors in $N$ also have to map somewhere in $\mathcal{R}$, and we need to detect that when such a vector maps to $\tau_i$, this does not indicate that $\mathbf{x}$ is a vector in $N$ labeled by $\tau_i$. This is the tricky part, and indeed we do not know how to accomplish this for completely general labeled linear matroids $(M, \sigma)$ and $(N, \tau)$. However, by imposing some structural restrictions on our matroids, we can still derive theorems of the same type that yield strong results when applied in the right way.

## 3.2 Canonical Functions for Labeled Matroids

Given a labeled matroid $(N, \tau)$ over vectors $\{\mathbf{w}_1, \ldots, \mathbf{w}_m\}$ in $\mathbb{F}^d$ and with pattern $\tau = \langle \tau_1, \ldots, \tau_m \rangle \in \mathcal{R}^m$, we construct the canonical function for $(N, \tau)$ as follows.

Let $J \subseteq [n]$ be any fixed subset of size $d$. We think of $J$ as the coordinates for the part of $\mathbf{x} \in \mathbb{F}^n$ that will correspond to the indicator function for $N$. (For now, the reader can fix this set to be $\{1, \ldots, d\}$ for simplicity.) For $J = \{j_1, \ldots, j_{d-1}\}$ and a vector $\mathbf{x} \in \mathbb{F}^n$, we will write $\mathbf{x}[J]$ to denote the vector $\mathbf{x}$ projected to its coordinates in $J$, i.e., $\mathbf{x}[J] = \{x_{j_1}, x_{j_2}, \ldots, x_{j_d}\}$. Below, we will write $\mathbf{x} = \{\mathbf{y}|\mathbf{z}\}$ to denote the decomposition of $\mathbf{x} \in \mathbb{F}^n$ into $\mathbf{y} = \mathbf{x}[J] \in \mathbb{F}^d$ and $\mathbf{z} = \mathbf{x} \setminus \mathbf{x}[J] \in \mathbb{F}^{n-d}$. We let $S$ be any subspace of $\mathbb{F}^{n-d}$ of high dimension. To be concrete, let us set the dimension to $n - d - 1$, which means that $S$ contains half the points of $\mathbb{F}^{n-d}$. (However, to make it easier to follow the arguments below, the reader can think of $S$ as being all of $\mathbb{F}^{n-d}$.) We let $b \in \mathcal{R}$ denote a "padding value." Loosely speaking, we will let our canonical function evaluate to $b$ on points that do not correspond to vectors in $N$. The parameters $J$ and $S$ are not important for the proofs, so we will suppress them in our definition of the canonical function for $(N, \tau)$, and the same goes for the dimension $n$.

**Definition 3.4 (Matroid canonical function).** Let $N = \{\mathbf{w}_1, \ldots, \mathbf{w}_m\}$ be any linear matroid in $\mathbb{F}^d$ labeled by $\tau = \langle \tau_1, \ldots, \tau_m \rangle \in \mathcal{R}^m$, and let $n$ be a dimension parameter. Fix any $J$ and $S$ as described above and write $\mathbf{x} \in \mathbb{F}^n$ as $\mathbf{x} = \{\mathbf{y}|\mathbf{z}\}$ for $\mathbf{y} = \mathbf{x}[J] \in \mathbb{F}^d$ and $\mathbf{z} = \mathbf{x} \setminus \mathbf{x}[J] \in \mathbb{F}^{n-d}$. Then for $b \in \mathcal{R}$, the *b-canonical function* $f^b_{N,\tau} : \mathbb{F}^n \to \mathcal{R}$ of the labeled matroid $(N, \tau)$ is defined by

$$f^b_{N,\tau}(x) = f^b_{N,\tau}(\{\mathbf{y}|\mathbf{z}\}) = \begin{cases} 0 & \text{if } \mathbf{y} = \mathbf{0} \text{ and } \mathbf{z} \in S; \\ \tau_j & \text{if } \mathbf{y} = \mathbf{w}_j \in N \text{ and } \mathbf{z} \in S; \\ b & \text{otherwise.} \end{cases}$$

The function $f^b_{N,\tau}$ encodes $(N, \tau)$ in the sense that it is dense in the matroid pattern $(N, \tau)$, as well as in any $(M, \sigma)$ that maps homomorphically into $(N, \tau)$.

**Lemma 3.5.** *If there is a labeled matroid homomorphism from $(M, \sigma)$ to $(N, \tau)$, then the function $f^b_{N,\tau} : \mathbb{F}^n \to \mathcal{R}$ is $\delta$-far from being $(M, \sigma)$-free, where $\delta > 0$ is some constant independent of $n$ and $b$.*

*Proof.* Let $\phi : (M, \sigma) \to (N, \tau)$ be a homomorphism. Suppose that $N = \{\mathbf{w}_1, \ldots, \mathbf{w}_m\}$ for $\mathbf{w}_j \in \mathbb{F}^d$, and let $L : \mathbb{F}^d \to \mathbb{F}^n$ be any linear transformation sending $\mathbf{w}_j$ to $\{\mathbf{w_j}|\mathbf{z_j}\}$ for arbitrary $\mathbf{z}_j \in S$. Then for all $\mathbf{v}_i \in M$, if $\phi(\mathbf{v}_i) = \mathbf{w}_j$ it is easy to verify that $f^b_{N,\tau}\big((L \circ \phi)(\mathbf{v}_i)\big) = f^b_{N,\tau}\big((L(\mathbf{w}_j)\big) = \tau_j = \sigma_i$, where the last equality holds since $\phi$ preserves labels. Hence, $f^b_{N,\tau}$ contains $(M, \sigma)$ at $L \circ \phi$.

The proof of $\delta$-farness closely follows a similar argument in [BCSX09]. Set $\delta = 1/(q|\mathbb{F}|^d)$ where $q = |\mathbb{F}| \geq 2$. Our approach is to show that any function that is $1/(q|\mathbb{F}|^d)$-close to $f^b_{N,\tau}$ contains $(M, \sigma)$ somewhere, which is clearly equivalent to that $f^b_{N,\tau}$ is $\delta$-far from $(M, \sigma)$-free. To this end, we fix a function $g$ with $\text{dist}(f^b_{N,\tau}, g) = \delta' < 1/(q|\mathbb{F}|^d)$. We will show that $g$ contains $(M, \sigma)$ at some linear map $L'$. Let $S$ be a subspace of $\mathbb{F}^{n-d}$ of codimension 1 as defined above. Because $|S| = |\mathbb{F}|^{n-d}/q$, clearly we have $\Pr_{\mathbf{y} \in \mathbb{F}^d, \mathbf{z} \in S}[f^b_{N,\tau}(\{\mathbf{y}|\mathbf{z}\}) \neq g(\{\mathbf{y}|\mathbf{z}\})] \leq q\delta'$. For $i \in [m]$, let $\delta_i = \Pr_{\mathbf{z} \in S}[f^b_{N,\tau}(\{\mathbf{w}_i|\mathbf{z}\}) \neq g(\{\mathbf{w}_i|\mathbf{z}\})]$. Since

$$\frac{1}{|\mathbb{F}|^d} \sum_{i=1}^{m} \delta_i \leq \Pr_{\mathbf{y} \in \mathbb{F}^d, \mathbf{z} \in S}\big[f^b_{N,\tau}(\{\mathbf{y}|\mathbf{z}\}) \neq g(\{\mathbf{y}|\mathbf{z}\})\big] \leq q\delta' , \tag{3.1}$$

we therefore have $\sum_{i=1}^{m} \delta_i \leq q|\mathbb{F}|^d \cdot \delta' < 1$. Now consider a random linear map $\tilde{L}_1 : \mathbb{F}^d \to S$, and its extension $\tilde{L} : \mathbb{F}^d \to \mathbb{F}^n$ given by $\tilde{L}(\mathbf{y}) = \{\mathbf{y}|\tilde{L}_1(\mathbf{y})\}$. For every non-zero $\mathbf{y}$ and in particular for $\mathbf{y} \in N$, we have that $\tilde{L}_1(\mathbf{y})$ is distributed uniformly over the subspace $S$. Thus, for any fixed $i \in [m]$, we have

$$\Pr_{\tilde{L}_1}[g(\tilde{L}(\mathbf{w}_i)) \neq \tau_i] = \Pr_{\tilde{L}_1}[g(\tilde{L}(\mathbf{w}_i)) \neq f^b_{N,\tau}(\tilde{L}(\mathbf{w}_i))] \leq \delta_i . \tag{3.2}$$

By the union bound, we get that

$$\Pr_{L_1}[\exists i \text{ such that } g(\tilde{L}(\mathbf{w}_i)) \neq \tau_i] \leq \sum_i \delta_i < 1 \ . \quad (3.3)$$

In other words, there exists a linear map $\tilde{L}_1$ (and thus $\tilde{L}$) such that for every $i$, $g(\tilde{L}(\mathbf{w}_i)) = \tau_i$ and so $g$ contains $(N, \tau)$ at $\tilde{L}$ and hence $(M, \sigma)$ at the linear map $L' = \tilde{L} \circ \phi$. □

The following observation will come in handy later on. Suppose that, instead of letting $S$ denote a fixed subspace as above, we associate to each $\mathbf{w} \in N \cup \{\mathbf{0}\}$ an independently chosen random subset $S_\mathbf{w} \subseteq \mathbb{F}^{n-d}$ of density $1/2$. And then, suppose that we modify the construction of the $b$-canonical function in Definition 3.4 to be:

$$f^b_{N,\tau}(x) = f^b_{N,\tau}(\{\mathbf{y}|\mathbf{z}\}) = \begin{cases} 0 & \text{if } \mathbf{y} = \mathbf{0} \text{ and } \mathbf{z} \in S_\mathbf{0}; \\ \tau_j & \text{if } \mathbf{y} = \mathbf{w}_j \in N \text{ and } \mathbf{z} \in S_{\mathbf{w}_j}; \\ b & \text{otherwise.} \end{cases} \quad (3.4)$$

We claim that Lemma 3.5 holds true with constant probability over the choices of all $S_\mathbf{w}$. This is because we can carry out the analysis of the last paragraph in the above proof with probability over $\tilde{L}_1$ and all $S_\mathbf{w}$ (instead of just $\tilde{L}_1$) and then apply a version of the Markov inequality. It is also easy to see that if $\tau_j$ is distinct from the padding value $b$, then different choices of $S_{\mathbf{w}_j}$ in (3.4) yield distinct functions. In this way, we can obtain a very large family of canonical functions that are $(N, \tau)$-free but far from being $(M, \sigma)$-free.

### 3.3 Two Dichotomy Theorems

In Section 3.2, we carried out the first two steps in the proof outline in Section 3.1. We now present two classes of pairs of labeled matroids $(M, \sigma)$ and $(N, \tau)$ for which we can also successfully complete the crucial third step, and thus establish a dichotomy in the sense that if containment does not hold between the two properties, then they must be well separated.

**Theorem 3.6 (First dichotomy theorem).** *Let $M, N$ be any linear matroids and let $\tau$ be any pattern for $N$. Then $(M, 1^*)$-freeness is contained in $(N, \tau)$-freeness if and only if there is a labeled matroid homomorphism from $(M, 1^*)$ to $(N, \tau)$; otherwise $(M, 1^*)$-freeness is $\delta$-separated from $(N, \tau)$-freeness.*

Before proving this theorem, we note that as a corollary it immediately yields our first main result in Theorem 1.3 on page 4 providing a full characterization of monotone matroid freeness properties. This follows simply by setting $\tau = 1^*$.

*Proof of Theorem 3.6.* The "if" part of the claim is Lemma 3.1. For the "only if" direction, let us assume that $(M, 1^*)$-freeness is contained in $(N, \tau)$-freeness. We need to prove that this implies that there is a labeled matroid homomorphism from $(M, 1^*)$ to $(N, \tau)$. Consider the canonical function $f^0_{N,\tau}$ for $(N, \tau)$ padded with zeros. We know from Lemma 3.5 that $f^0_{N,\tau}$ is not $(N, \tau)$-free. Hence if $(M, 1^*)$-freeness is contained in $(N, \tau)$-freeness it cannot be $(M, 1^*)$-free either, so suppose it contains $(M, 1^*)$ at the linear transformation $L : M \to \mathbb{F}^n$. We claim that if we let $\pi$ be the projection that maps $x = \{\mathbf{y}|\mathbf{z}\}$ to $\mathbf{y}$, then $\pi \circ L$ must be a labeled matroid homomorphism from $(M, 1^*)$ to $(N, \tau)$.

To see this, note first that the map $\pi \circ L$ is clearly linear. We need to check that it sends every vector $\mathbf{v}_i \in M$ to some vector $\mathbf{w}_j$ in $N$ and in addition that the labels of the vectors are preserved. But since $M$ has the monotone pattern $\langle 1^*\rangle$ the label is always 1; hence, by assumption we have $f^0_{N,\tau}(L(\mathbf{v}_i)) = 1$ for all $i$. It follows from the way the canonical function was constructed in Definition 3.4 that we must have $L(\mathbf{v}_i) = \{\mathbf{y}_i|\mathbf{z}_i\}$ where $\mathbf{y}_i = \mathbf{w}_j$ for some $\mathbf{w}_j \in N$ labeled by $\tau_j = 1$, since these are the only vectors for which $f^0_{N,\tau}$ evaluates to 1. Thus, $\pi \circ L$ is a labeled matroid homomorphism from $(M, 1^*)$ to $(N, \tau)$, which establishes the claim. □

For future use, we note that the choice of $S$ in the construction of the canonical function $f^0_{N,\tau}$ did not matter at all in the above proof. This fact will allow us later to exploit the observation made at the end of Section 3.2.

We can use Theorem 3.6 not only to separate monotone properties from non-monotone ones, but also to separate two non-monotone properties from each other. Namely, let $(N^1, \tau^1)$ and $(N^2, \tau^2)$ be non-monotone labeled matroids such that $(N^1, \tau^1)$ is a submatroid of $(N^2, \tau^2)$ (for instance, by being a labeled subgraph). Suppose furthermore that we can find a monotone matroid $(M, 1^*)$ having the property that $(M, 1^*) \hookrightarrow (N^2, \tau^2)$ but $(M, 1^*) \not\hookrightarrow (N^1, \tau^1)$. Then it follows from Lemma 3.1 and Theorem 3.6 that $(N^1, \tau^1)$-freeness must be strictly contained in $(N^2, \tau^2)$-freeness. We will see examples of such results in Section 5.

However, while this is already considerably stronger than the monotone separation results in [BCSX09], it is still not quite satisfactory. The problem is that results obtained in this manner do not show that non-monotonicity adds anything essential to matroid freeness properties. For all that we know, it might be the case that $(N^1, \tau^1)$-freeness is identical to $(M, 1^*)$-freeness so that the only essential constraint is the monotone one and the non-monotone constraints are just syntactic sugar. Our second dichotomy theorem, while being more restricted in the structural conditions it places on the matroids, is also much more powerful in that it directly separates non-monotone properties without going via monotone ones.

**Theorem 3.7 (Second dichotomy theorem).** *Let $N$ be a matroid in $\mathbb{F}_2^d$ in standard representation containing $M(K_{d'+1})$ as a submatroid on the first $d' \leq d$ basis vectors (i.e., for $\mathbf{e}_1, \ldots, \mathbf{e}_{d'}$ all sums $\mathbf{e}_i + \mathbf{e}_j$, $1 \leq i < j \leq d'$ are also vectors in $N$), and let $0^{d'}1^*$ be the pattern for $N$ that gives $0$-labels to the vectors $\mathbf{e}_1, \ldots, \mathbf{e}_{d'}$ and $1$-labels to all other vectors.*

*Then $(M(K_c), 0^{c-1}1^*)$-freeness is contained in $(N, 0^{d'}1^*)$-freeness if and only if there is a labeled matroid homomorphism from $(M(K_c), 0^{c-1}1^*)$ to $(N, 0^{d'}1^*)$; otherwise $(M(K_c), 0^{c-1}1^*)$-freeness is $\delta$-separated from $(N, 0^{d'}1^*)$-freeness.*

The two theorems 3.6 and 3.7 together constitute the formal version of our second main result in Theorem 1.4. We remark that in contrast to Theorem 3.6, in Theorem 3.7 we are only considering the case when the range of the functions in our properties is $\mathcal{R} = \{0, 1\}$.

*Proof of Theorem 3.7.* The "if" direction is again Lemma 3.1. For the "only if" direction, suppose that $(M(K_c), 0^{c-1}1^*)$-freeness is contained in $(N, 0^{d'}1^*)$-freeness. Consider $f = f^1_{N, 0^{d'}1^*}$, where we point out that we are now padding the canonical function with ones (as opposed to the zero-padding in the proof of Theorem 3.6). We know from Lemma 3.5 that $f$ is far from being $(N, 0^{d'}1^*)$-free. Hence, by our assumption it cannot be $(M(K_c), 0^{c-1}1^*)$-free either. Suppose that $f$ contains $(M(K_c), 0^{c-1}1^*)$ at $L : M(K_c) \to \mathbb{F}^n$. Let $\pi$ be the projection that maps $\mathbf{x} = \{\mathbf{y}|\mathbf{z}\}$ to $\mathbf{y}$. We want to argue that $\pi \circ L$ must be a labeled matroid homomorphism from $(M(K_c), 0^{c-1}1^*)$ to $(N, 0^{d'}1^*)$.

Let us first focus on the basis vectors in $M(K_c)$, which we will denote $\mathbf{f}_1, \ldots, \mathbf{f}_{c-1}$, and which are all $0$-labeled. Since $f$ applied on the image of $M(K_c)$ under $L$ evaluates to the pattern $\langle 0^{c-1}1^* \rangle$, we have $f(L(\mathbf{f}_i)) = 0$ for all $i \in [c-1]$. Looking at the definition of $f$, this means that $L(\mathbf{f}_i) = \{\mathbf{y}_i|\mathbf{z}_i\}$ where either $\mathbf{y}_i = \mathbf{e}_l$ for some $\mathbf{e}_l \in N$, $l \leq d'$, or else $\mathbf{y}_i = \mathbf{0}$. We also note for the record, since we will need it later in the proof, that we must have $\mathbf{z}_i \in S$ in both of these cases.

Clearly, if $L(\mathbf{f}_i) = \{\mathbf{y}_i|\mathbf{z}_i\}$ for $\mathbf{y}_i = \mathbf{0}$, the linear map $\pi \circ L$ is no matroid homomorphism (since $\mathbf{0} \notin N$) and the construction breaks down. We claim, however, that this can never happen. Given this claim, all basis vectors $\mathbf{f}_i \in M(K_c)$ must then be mapped by $L$ to $\{\mathbf{e}_{l_i}|\mathbf{z}_i\}$ for some $\mathbf{e}_{l_i} \in N$, $l_i \leq d'$ and some $\mathbf{z}_i \in S$. The only other vectors in $M(K_c)$ are sums $\mathbf{f}_i + \mathbf{f}_j$, and by linearity we have $(\pi \circ L)(\mathbf{f}_i + \mathbf{f}_j) = \mathbf{e}_{l_i} + \mathbf{e}_{l_j} \in \mathbb{F}^d$ for $l_i, l_j \leq d'$. Again we have two cases. If $\mathbf{e}_{l_i} \neq \mathbf{e}_{l_j}$, then by the assumptions in the statement of the theorem we have that $\mathbf{e}_{l_i} + \mathbf{e}_{l_j}$ is a vector in $N$ labeled by $1$ as desired. If, however, $\mathbf{e}_{l_i} = \mathbf{e}_{l_j}$, then $\mathbf{f}_i + \mathbf{f}_j$ gets mapped to $\mathbf{0}$ and the construction breaks down. This cannot happen, however, since it would imply that

SEPARATIONS OF MATROID FREENESS PROPERTIES

$f(L(\mathbf{f}_i + \mathbf{f}_j)) = f(\{\mathbf{0}|\mathbf{z}_i + \mathbf{z}_j\}) = 0$. (Again for the record, this holds because $\mathbf{z}_i, \mathbf{z}_j \in S$ implies that also $\mathbf{z}_i + \mathbf{z}_j \in S$, since $S$ is a linear subspace). But $f(L(\mathbf{f}_i + \mathbf{f}_j)) = 0 \neq 1$ contradicts the assumption that $f$ evaluates to the pattern $\langle 0^{c-1}1^* \rangle$ on the image of $M(K_c)$ under $L$. Hence, $\pi \circ L$ maps $M(K_c)$ into $N$ while preserving labels, i.e., it is a labeled matroid homomorphism.

It remains to prove the claim that $L(\mathbf{f}_i) \neq \{\mathbf{0}|\mathbf{z}_i\}$ for all basis vectors $\mathbf{f}_i \in M(K_c)$. Suppose on the contrary that there is a vector $\mathbf{f}_i$ such that $L(\mathbf{f}_i) = \{\mathbf{0}|\mathbf{z}_i\}$. Fix some other basis vector $\mathbf{f}_j$, $j \neq i$, in $M(K_c)$ that is mapped by $L$ to $\{\mathbf{y}_j|\mathbf{z}_j\}$ for $\mathbf{y}_j \in \{\mathbf{e}_1, \ldots, \mathbf{e}_{d'}\} \cup \{\mathbf{0}\}$, and consider $L(\mathbf{f}_i + \mathbf{f}_j) = L(\mathbf{f}_i) + L(\mathbf{f}_j) = \{\mathbf{y}_j|\mathbf{z}_i + \mathbf{z}_j\}$. By assumption $f(L(\mathbf{f}_i)) = f(L(\mathbf{f}_j)) = 0$, which by the definition of $f$ implies that $\mathbf{z}_i, \mathbf{z}_j \in S$. This in turn means that $f(L(\mathbf{f}_i + \mathbf{f}_j)) = f(L(\mathbf{f}_j)) = 0$. But this is again a contradiction to the assumption that $f$ evaluates to the pattern $\langle 0^{c-1}1^* \rangle$, which requires that $f(L(\mathbf{f}_i + \mathbf{f}_j)) = 1$. The claim follows, and the proof of the theorem is complete. □

We remark is that unlike the proof for Theorem 3.6, here we crucially use the fact that $S$ is a linear subspace and so we cannot replace $S$ by, for instance, a random subset.

## 4 Some Labeled Graphic Matroid Non-Homomorphisms

In order for the method developed in the previous section to be useful, we need to find (families of) labeled matroids that do not embed homomorphically into each other. In this section, we establish such matroid non-homomorphism results for graphic matroids. Recall that for labeled graphic matroids $(M(G), \sigma)$ and $(M(H), \tau)$, which we will from now identify with their underlying labeled graphs $(G, \sigma)$ and $(H, \tau)$ for ease of notation, the matroid vectors correspond to edges in the graphs, and a labeled matroid homomorphism is a mapping of edges to edges that preserves labels and cycles.

The key to all of our non-homomorphism results is (the proof of) the following lemma.

**Lemma 4.1.** *For all $d \geq 5$, there is no labeled matroid homomorphism from $(K_d, \sigma)$ to $(K_{d-1}, \tau)$ for any patterns $\sigma$ and $\tau$.*

We remark that as shown in Proposition 3.3, there is in fact a homomorphism from $(K_4, 1^*)$ to $(K_3, 1^*)$. Thus, the condition $d \geq 5$ above is necessary.

To prove Lemma 4.1, we ignore the patterns $\sigma$ and $\tau$ and instead argue directly that regardless of what these patterns look like, there can exist no edge homomorphism from $K_d$ to $K_{d-1}$ that preserves cycle structure. This argument rests on two simple but very useful claims.

To state these claims, we recall from Definition 2.6 that the standard representation of $K_d$ is to fix some vertex $v$ and pick as a basis the vectors corresponding to all edges $e$ incident to this vertex. Even once we have fixed such a vertex and associated its edges with unit vectors $\mathbf{e}_1, \mathbf{e}_2, \ldots, \mathbf{e}_{d-1}$, we can get another essentially equivalent basis by fixing any other vertex and looking at the vectors corresponding to edges incident to that vertex instead, as explained after Definition 2.6. We will refer to any such basis, which in this section we identify with the corresponding set of edges, as a *standard form basis*.

**Claim 4.2.** *For any $c, d \geq 3$, if $(K_d, \sigma) \hookrightarrow (K_c, \tau)$, then it holds that any two incident edges in $K_d$ must map to distinct edges in $K_c$.*

**Claim 4.3.** *For any $d \geq 5$ and $c \geq 4$, if $(K_d, \sigma) \hookrightarrow (K_c, \tau)$, then all edges in any standard form basis of $K_d$ must map to edges in $K_c$ that are all incident to one common vertex.*

Given these two claims, Lemma 4.1 follows immediately by a pigeonhole argument: since a standard form basis in $K_d$ has $d - 1$ edges while all vertices in $K_{d-1}$ only has $d - 2$ incident edges, the claims 4.2 and 4.3 cannot possibly both hold simultaneously. This proves the lemma by contradiction.

It remains to establish the claims, and we do so next. Note again that we write $v$ to denote a vertex and $e$ to denote an edge, whereas vectors are denoted $\mathbf{e}, \mathbf{f}, \mathbf{v}, \mathbf{w}$, et cetera. In what follows below, we will go freely back and forth between edge and vector representation of the graphic matroids.

The first claim is more or less immediate.

*Proof of Claim 4.2.* Let $e_1 = (v_i, v_j)$ and $e_2 = (v_i, v_k)$ be two incident edges in $K_d$ and suppose that they map to the same edge in $K_c$. Now $e_1$ and $e_2$ form a cycle in $K_d$ together with $e_3 = (v_j, v_k)$, but since $K_c$ does not have self-loops there is no way to map $e_3$ to an edge in $K_c$ so that this cycle is preserved. (Or, reasoning in terms of matroid vectors in a linear space over $\mathbb{F}_2$, since the vectors for $e_1$ and $e_3$ map to the same vector in $K_c$ and hence cancel, there is no way to map the third vector to a non-zero vector in $K_3$ so that the sum of the images of all three vectors cancel.) □

The second claim is not much harder, but requires a little more work.

*Proof of Claim 4.3.* Suppose $K_d$ embeds into $K_c$ by a linear map $\phi$ and let $\{\mathbf{f}_1, \ldots, \mathbf{f}_{d-1}\}$ be the basis vectors of $K_d$ and $\{\mathbf{e}_1, \ldots, \mathbf{e}_{c-1}\}$ be the basis vectors of $K_c$. We show by induction on $k$, $1 \le k \le d-1$, that $\{\mathbf{f}_1, \ldots, \mathbf{f}_k\}$ must map to (distinct) edges incident to some common vertex $v$ in $K_c$.

Without loss of generality, assume $\phi(\mathbf{f}_1) = \mathbf{e}_1$ (if $\mathbf{f}_1$ would map to any other vector we just make a basis change in $K_c$ as explained after Definition 2.6). By Claim 4.2 we have $\phi(\mathbf{f}_2) \ne \mathbf{e}_1$. Also note that $\phi(\mathbf{f}_2) \ne \mathbf{e}_i + \mathbf{e}_j$ with $1 < i < j$, because this would imply $\phi(\mathbf{f}_1 + \mathbf{f}_2) = \mathbf{e}_1 + \mathbf{e}_i + \mathbf{e}_j$, which is weight-3 vector that is not a member of $K_c$. (In terms of edges, this would correspond to incident edges $e_1 = (v_i, v_j)$ and $e_2 = (v_i, v_k)$ in $K_d$ mapping to non-incident edges in $K_c$, but if so there would be no way the map $\phi$ could preserve the cycle of $e_1$ and $e_2$ with $e_3 = (v_j, v_k)$.)

Therefore we are left with two cases: $\phi(\mathbf{f}_2) = \mathbf{e}_i$ or $\phi(\mathbf{f}_2) = \mathbf{e}_1 + \mathbf{e}_i$ for some $i > 1$, and because of symmetry we may choose $i = 2$ without loss of generality. Let us analyze these two cases.[5]

**Case 1** ($\phi(\mathbf{f}_2) = \mathbf{e}_2$): Consider where $\phi$ can send $\mathbf{f}_3$. Clearly $\phi(\mathbf{f}_3) \notin \{\mathbf{e}_1, \mathbf{e}_2\} = \{\phi(\mathbf{f}_1), \phi(\mathbf{f}_2)\} =$ by the distinctness in Claim 4.2. Also, we claim that $\phi(\mathbf{f}_3) \ne \mathbf{e}_1 + \mathbf{e}_2$. To see this, observe that if $\{\phi(\mathbf{f}_1), \phi(\mathbf{f}_2), \phi(\mathbf{f}_3)\} = \{\mathbf{e}_1, \mathbf{e}_2, \mathbf{e}_1 + \mathbf{e}_2\}$, then $\phi$ would have to map $\mathbf{f}_4$ to some vector outside of this set by distinctness, but if so it in turns follows that $\phi(\mathbf{f}_3 + \mathbf{f}_4) = \phi(\mathbf{f}_3) + \phi(\mathbf{f}_4)$ would be a vector of weight at least 3. (Notice that here we crucially use $d \ge 5$). To conclude, we argue that in fact $\phi(\mathbf{f}_3) \ne \mathbf{e}_i + \mathbf{e}_j$ for any $i < j$ with $j > 2$. For if $i = 1$, say, it follows just as above that $\phi(\mathbf{f}_2 + \mathbf{f}_2) = \phi(\mathbf{f}_2) + \phi(\mathbf{f}_3)$ would be a weight-3 vector. Hence we must have $\phi(\mathbf{f}_3) = \mathbf{e}_i$ for some $i$, which we may without loss of generality set to $i = 3$.

**Case 2** ($\phi(\mathbf{f}_2) = \mathbf{e}_1 + \mathbf{e}_2$): In this case we have $\phi(\mathbf{f}_3) \ne \mathbf{e}_1$ by distinctness, and for the same reasons as in case 1 we deduce that $\phi(\mathbf{f}_3) \ne \mathbf{e}_2$. Furthermore, $\phi(\mathbf{f}_3) \ne \mathbf{e}_i$ for $i > 2$, since if so $\phi(\mathbf{f}_2 + \mathbf{f}_3)$ would be a weight-3 vector. Hence, $\phi(\mathbf{f}_3) = \mathbf{e}_i + \mathbf{e}_j$ for some $i < j$. But then we must have $i = 1$, since otherwise $\phi(\mathbf{f}_1 + \mathbf{f}_3)$ would have weight 3. We have proven that in this case as well, the vectors $\phi(\mathbf{f}_1) = \mathbf{e}_1$, $\phi(\mathbf{f}_2) = \mathbf{e}_1 + \mathbf{e}_2$, and $\phi(\mathbf{f}_3) = \mathbf{e}_1 + \mathbf{e}_j$ must correspond to edges incident to a common vertex.

Now we proceed to the inductive step. Suppose $\phi$ maps the vectors $\{\mathbf{f}_1, \ldots, \mathbf{f}_k\}$ to $k$ edges incident to a single vertex in $K_c$ for some $k \ge 4$. Without loss of generality, we assume that the vectors are mapped to $\mathbf{e}_1, \ldots, \mathbf{e}_k$ in $K_c$ (because again, since we know that the edges are incident we are free to make a basis change in $K_c$ so that we get the standard basis in terms of unit vectors). Consider the image of edge $\mathbf{f}_{k+1}$ in $K_c$. Clearly, $\phi(\mathbf{f}_{k+1}) \neq \mathbf{e}_i$ for any $i \in [k]$ by distinctness. It is also easy to see that $\phi(\mathbf{f}_{k+1})$ cannot be any weight-2 vector. For if we would have $\phi(\mathbf{f}_{k+1}) = \mathbf{e}_i + \mathbf{e}_j$, then since $k \geq 3$ there would exist some $l \in [k]$ with $l \notin \{i, j\}$ such that $\phi(\mathbf{f}_l + \mathbf{f}_{k+1}) = \mathbf{e}_i + \mathbf{e}_j + \mathbf{e}_l \notin K_c$, which is a contradiction. Therefore, $\phi(\mathbf{f}_{k+1}) = \mathbf{e}_i$ for some $i > k$. This completes the induction step and thus finishes the proof of the claim. □

---

[5]In fact, the attentive reader might have noted here that without loss of generality we can restrict ourselves to only one case and fix $\phi(\mathbf{f}_2) = \mathbf{e}_2$. This is so since in the other case we can again make a basis change in $K_c$ as described after Definition 2.6 to get a new standard basis containing $\mathbf{e}_1$ and $\mathbf{e}_1 + \mathbf{e}_2$. However, we believe that a formal case analysis as given in this proof, although strictly speaking unnecessary, is easier to follow.

Next, we study labeled homomorphisms from $K_d$ into itself, and rule out that a monotone pattern can be mapped homomorphically into a non-monotone one.

**Lemma 4.4.** *For all $d \geq 5$ and $c \geq 1$, there is no labeled matroid homomorphism from $(K_d, 1^*)$ to $(K_d, 0^c 1^*)$.*

*Proof.* By Claims 4.2 and 4.3 it must be the case that all the basis vectors of $K_d$ map in a one-to-one and onto fashion to distinct edges adjacent to a single vertex $v_1$ in $K_d$, say $\phi(\mathbf{e}_1) = (v_1, v_2), \phi(\mathbf{e}_2) = (v_1, v_3), \ldots, \phi(\mathbf{e}_{d-1}) = (v_1, v_{d-1})$. Since labels are preserved by the mapping, none of the basis vectors $\mathbf{e}_i$ maps to a 0-labeled edge in $(K_d, 0^c 1^*)$. However, consider any 0-labeled edge $e = (v_i, v_j)$ in $(K_d, 0^c 1^*)$. Notice that since $\phi$ is a homomorphism we must have $\phi(\mathbf{e}_i + \mathbf{e}_j) = \phi(\mathbf{e}_i) + \phi(\mathbf{e}_j) = (v_i, v_j)$, that is, $\mathbf{e}_i + \mathbf{e}_j$ in $(K_d, 1^*)$ maps to this 0-labeled edge. But if so $\phi$ is not label-preserving, since all vectors in $(K_d, 1^*)$ are labeled 1. Contradiction. □

A second useful lemma about non-monotone patterns is as follows.

**Lemma 4.5.** *For all $d \geq 4$, there is no labeled matroid homomorphism from $(K_d, 0^{d-1} 1^*)$ to $(K_d, 0^{d-2} 1^*)$.*

*Proof.* By Claims 4.2 and 4.3 it must be that all the $d - 1$ basis vectors of $(K_d, 0^{d-1} 1^*)$ map in a one-to-one fashion to distinct edges adjacent to a single vertex in $(K_d, 0^{d-1} 1^*)$. Notice that the basis vectors of $(K_d, 0^{d-1} 1^*)$ are all labeled by 0 and hence they have to map to the 0-labeled edges in $(K_d, 0^{d-2} 1^*)$. But there are only $d - 2$ such edges in $(K_d, 0^{d-2} 1^*)$, and since the basis vectors must be mapped to distinct edges it immediately follows that such a homomorphism is not possible to construct. □

## 5 Infinite Hierarchies of Well Separated Matroid Freeness Properties

We have finally reached the point where we can put all the material in Sections 3 and 4 together and prove the existence of infinite hierarchies of $(M, \sigma)$-freeness properties as claimed in the introduction. We remark that the techniques we have developed could be used to yield many different such hierarchies, but for brevity and concreteness we will focus below on one particular result that illustrates this general point. Namely, the lemmas proven in this section all lead up to Theorem 5.4, which is the formal statement of our third main result claimed in Theorem 1.5 in the introduction.

Let us start by proving that monotone $M(K_d)$-freeness properties form a strict hierarchy. Note that we will continue the mild abuse of notation introduced in Section 4 by identifying a graphic matroid $(M(G), \sigma)$ and its underlying labeled graph $(G, \sigma)$.

**Lemma 5.1.** *For $d \geq 4$, the $(K_d, 1^*)$-freeness properties form an infinite hierarchy of strictly contained properties.*

*Proof.* Since $(K_d, 1^*)$ is a labeled subgraph of $(K_{d+1}, 1^*)$ we have that $(K_d, 1^*)$-freeness is contained in $(K_{d+1}, 1^*)$-freeness by Corollary 3.2. Since there is no homomorphism from $(K_{d+1}, 1^*)$ to $(K_d, 1^*)$ for $d \geq 4$ according to Lemma 4.1, we conclude from Theorem 3.6 that the containment must be strict in a property testing sense. □

We remark that this lemma improves on a similar theorem of [BCSX09], which could only show separation between the graphic matroids of $K_d$ and $K_{\binom{d}{2}+2}$. But we can strengthen Lemma 5.1 even further as follows.

**Lemma 5.2.** *For $d \geq 4$ and any sequence $\{c_d\}_{d=4}^{\infty}$ with $1 \leq c_d < d$, the $(K_d, 0^{c_d}1^*)$-freeness properties form an infinite sequence of properties such that $(K_{d+1}, 0^{c_{d+1}}1^*)$-freeness is $\delta$-separated from $(K_d, 0^{c_d}1^*)$-freeness. If in addition the $c_d$-sequence is monotone increasing, i.e., $c_{d+1} \geq c_d$, then we get an infinite hierarchy of strictly contained properties.*

*Proof.* $(K_d, 1^*)$ does not embed in $(K_d, 0^{c_d}1^*)$ by Lemma 4.4. However, $(K_d, 1^*)$ is a labeled subgraph of $(K_{d+1}, 0^{c_{d+1}}1^*)$, since if we throw away the unique vertex in $(K_{d+1}, 0^{c_{d+1}}1^*)$ incident to all 0-labeled edges what we have left is exactly $(K_d, 1^*)$. It follows that the function $f^0_{K_d, 0^{c_d}1^*}$, which is far from being $(K_d, 0^{c_d}1^*)$-free, is $(K_d, 1^*)$-free by Theorem 3.6 and hence $(K_{d+1}, 0^{c_{d+1}}1^*)$-free by Corollary 3.2

If it furthermore holds that $c_d \leq c_{d+1} \leq d$, then $(K_d, 0^{c_d}1^*)$ is a labeled subgraph of $(K_{d+1}, 0^{c_{d+1}}1^*)$, which gives containment of the corresponding matroid freeness properties (again by Corollary 3.2). □

Observe that Lemma 5.2 is indeed a strengthening of Lemma 5.1. This is so since the functions $f^0_{K_d, 0^{c_d}1^*}$, which are $(K_d, 1^*)$-free but far from $(K_{d-1}, 1^*)$-free, witness that $(K_d, 1^*)$-freeness is $\delta$-separated from $(K_{d-1}, 1^*)$-freeness (and containment in the other direction is obvious since $(K_{d-1}, 1^*)$ is a subgraph of $(K_d, 1^*)$).

Notice, however, that as discussed after the proof of Theorem 3.6, the way we establish Lemma 5.2 does not ensure that non-monotone matroid freeness properties are nontrivial. We might worry that perhaps all the non-monotone properties coincide with the intermediate monotone properties of $(K_d, 1^*)$-freeness used to obtain the separation. The next lemma provides some assurance us by conclusively ruling out this possibility.

**Lemma 5.3.** *For $d \geq 4$, it holds that $(K_{d+1}, 0^d 1^*)$-freeness is $\delta$-separated from the union of $(K_d, 0^{d-1}1^*)$-freeness and $(K_d, 1^*)$-freeness.*

*Proof.* Note first that this lemma is conceptually different from the preceding ones, since here we need to find a function that is $(K_{d+1}, 0^d 1^*)$-free but is simultaneously far from being $(K_d, 0^{d-1}1^*)$-free and $(K_d, 1^*)$-free. On the face of it, such cases are not covered by the techniques in Sections 3 and 4, which only relates pair of labeled matroids. However, there is a way to get around this obstacle by finding an "intermediate" labeled matroid such that $(K_d, 0^{d-1}1^*)$ and $(K_d, 1^*)$ both embed into this matroid but $(K_{d+1}, 0^d 1^*)$ does not. We pick this intermediate matroid to be $(K_{d+1}, 0^{d-1}1^*)$, that is, the graphic matroid over the complete graph on $d+1$ vertices that have all edges *but one* in the standard basis labeled by 0 and has 1-labels everywhere else.

It is easy to check that $(K_d, 0^{d-1}1^*)$ and $(K_d, 1^*)$ are both labeled subgraphs of $(K_{d+1}, 0^{d-1}1^*)$. Consequently, we can appeal to Lemma 3.5 to conclude that the canonical function $f^1_{K_{d+1}, 0^{d-1}1^*}$ is dense in violations of both $(K_d, 0^{d-1}1^*)$-freeness and $(K_d, 1^*)$-freeness. However, Lemma 4.5 shows that $(K_{d+1}, 0^d 1^*)$ does not embed homomorphically into $(K_{d+1}, 0^{d-1}1^*)$, and therefore $f^1_{K_{d+1}, 0^{d-1}1^*}$ must be $(K_{d+1}, 0^d 1^*)$-free according to Theorem 3.7. The lemma follows. □

Combining all of these lemmas, we can prove that non-monotone graphic matroid freeness properties provide infinite hierarchies of strictly contained properties. The reader might be helped in parsing the next theorem and its proof by looking at the illustration in Figure 2.

**Theorem 5.4.** *Let $\mathcal{A}_d$ denote the set of all $(K_d, 1^*)$-free functions and $\mathcal{B}_d$ the set of all $(K_d, 0^{d-1}1^*)$-free functions in $\bigcup_{n \in \mathbb{N}^+} \{\mathbb{F}_2^n \to \{0,1\}\}$. Then the following holds:*

1. *For $d \geq 4$, $\mathcal{A}_d$ forms an infinite hierarchy of strictly contained properties.*

2. *For $d \geq 4$, $\mathcal{B}_d$ forms an infinite hierarchy of strictly contained properties.*

3. $\bigcap_{d=4}^{\infty} (\mathcal{A}_d \cap \mathcal{B}_d) \neq \emptyset$.

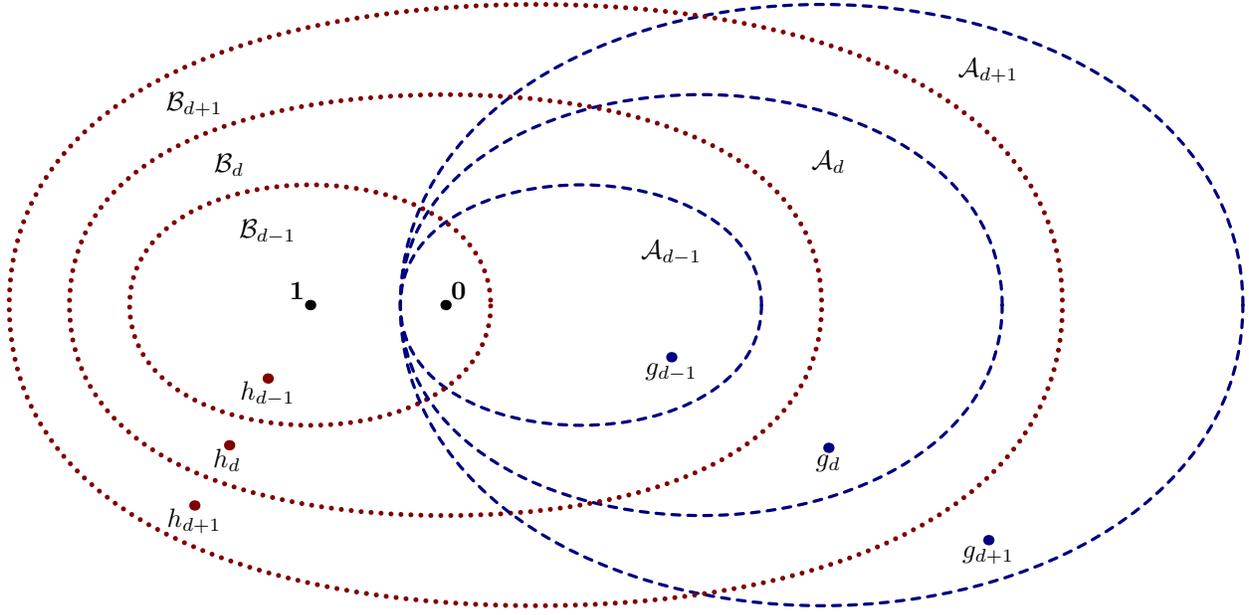

**Figure 2:** Illustration of hierarchies of properties $\mathcal{A}_d$ (dashed) and $\mathcal{B}_d$ (dotted) and separating functions.

4. $\left(\bigcup_{d=4}^{\infty} \mathcal{B}_d\right) \setminus \left(\bigcup_{d=4}^{\infty} \mathcal{A}_d\right) \neq \emptyset$, and in fact the former union of properties is $\delta$-separated from the latter.

5. For $d \geq 4$, $\mathcal{A}_d \cup \mathcal{B}_d$ is strictly contained in $\mathcal{B}_{d+1}$.

6. For $d \geq 4$, $\mathcal{A}_d$ and $\mathcal{B}_d$ are mutually well separated from one another.

7. For $d \geq 5$, all the properties $\mathcal{A}_d$, $\mathcal{B}_d$, $\mathcal{A}_d \setminus \mathcal{A}_{d-1}$, and $\mathcal{B}_d \setminus \mathcal{B}_{d-1}$ are far from being low-degree polynomials.

*Proof.* The proofs of Claims 1 and 2 were given in Lemma 5.1 and Lemma 5.2, respectively, and as was discussed after the proof of Lemma 5.2 we can in fact see that both of these hierarchies are witnessed by the functions $g_d = f^0_{K_d, 0^{c_d}1^*}$ as plotted schematically in Figure 2. To show claim 3, we observe that the constant function $\mathbf{0} : \mathbb{F}_2^n \to \{0, 1\}$ sending all points to 0 belongs to all of the properties $\mathcal{A}_d$ and $\mathcal{B}_d$. Claim 4 similarly follows since the constant function $\mathbf{1} : \mathbb{F}_2^n \to \{0, 1\}$ sending all points to 1 must be $(K_d, 0^{d-1}1^*)$-free simply by virtue of not having any zeros, while it is far from $(K_d, 1^*)$-free for exactly the same reason. Claim 5 was established in Lemma 5.3, using the functions $h_d = f^1_{K_d, 0^{d-2}1^*}$, and taken together, the functions $g_d$ and $h_d$ can be seen to witness the mutual separations in claim 6.

Consider finally claim 7. Recall that what we want to say is that the properties $\mathcal{A}_d$ and $\mathcal{B}_d$ are "new" properties not known to be testable before. Note that by necessity, such a statement must be somewhat informal — unless we can provide a full enumeration of all testable properties and separate our new properties from all of them via some kind of diagonalization argument, which arguably seems neither feasible nor particularly reasonable. But what seems natural to do is to prove formally that matroid freeness properties are not identical to the "usual suspects", which in this case would seem to be low-degree polynomials.

We remark that one can first make the easy observation that it cannot possibly be the case that *all* of the properties $\mathcal{A}_d$ and $\mathcal{B}_d$ are low-degree polynomials. If they were, there would be no way they could nest and intersect in the way shown in Figure 2, since low-degree polynomials just form one strict hierarchy of concentric circles with respect to degree. We want to prove something stronger, however, namely that *none* of the properties $\mathcal{A}_d$ and $\mathcal{B}_d$ can be just low-degree polynomials. The way we do this is to observe that we can modify the construction of canonical functions in Definition 3.4 slightly as in Equation (3.4) to get a huge family of canonical functions instead of just one, and that this can be done in such a way that Lemma 3.5 and Theorem 3.6 still hold (we refer to the discussion in Section 3 for the details). What this means is that we can think of every witnessing function $g_d$ in Figure 2 as being a large, dense cloud of such functions, and these functions are simply far too many to all be low-degree polynomials, or even to be close to low-degree polynomials. This establishes the claim, and the theorem follows. □

The attentive reader might have noticed that there is one natural piece missing in Theorem 5.4, namely the claim that $\mathcal{B}_{d+1} \setminus (\mathcal{A}_d \cup \mathcal{B}_d)$ is also far from being just low-degree polynomials. This seems very likely to be the case but we are currently unable to prove this. It is easy to prove that there are polynomials of very high degree in $\mathcal{B}_{d+1} \setminus (\mathcal{A}_d \cup \mathcal{B}_d)$. The way to see this is to define $h_d = f^1_{K_d, 0^{d-2}1^*}$ as in Definition 3.4 except that we choose $S$ to be a very small subspace, say of constant dimension. Then $h_d$ will evaluate to 1 everywhere except at a constant number of points, and therefore it cannot possibly be a low-degree polynomial. However, all such $h_d$ are also close to the constant function evaluating to 1 everywhere, which has very low degree indeed. One natural idea is instead to pick the set $S$ randomly to get a large number of functions $h_d$, and then argue that they are so many that here must be examples of such functions that are far from being low-degree polynomials. Unfortunately, this does not work. The proof of Theorem 3.7 turns out to be surprisingly delicate, and provably requires $S$ to be a subspace. We still strongly believe that the properties $\mathcal{B}_{d+1} \setminus (\mathcal{A}_d \cup \mathcal{B}_d)$ are far from low-degree polynomials for all $d \geq 4$, but it seems new techniques would be required to establish such a claim.

## 6   Concluding Remarks

Motivated by questions raised in [BCSX09] and the recent testability results in [BGS10], in this paper we have studied the semantics of matroid freeness properties, and in particular the problem of determining when two syntactically different matroid constraints in fact also encode *semantically* different properties. We have developed a new method for comparing matroid freeness constraints based on the concept of *labeled matroid homomorphisms*, and have shown that for a surprisingly broad class of matroid freeness properties this method exactly characterizes the relation between two matroid freeness properties. Even more, when the method works, it in fact establishes a strong dichotomy in the sense that either one property must be contained in the other or the properties are strictly distinct in a property testing sense. As a consequence, we established that results in [BGS10] do indeed provide infinite hierarchies of new properties not known to have been testable before.

Our work raises many interesting questions which we believe merit further study. Perhaps the most obvious open problem is in what generality our method of characterizing matroid freeness properties in terms of labeled homomorphisms can be made to work, and in particular whether it can be extended to arbitrary labeled graphic matroids, or even arbitrary linear matroids. That is, is it always true that $(M, \sigma)$-freeness is contained in $(N, \tau)$-freeness if and only if there is a labeled matroid homomorphism from $(M, \sigma)$ to $(N, \tau)$, and that the two properties must be well separated otherwise? As was explained in Section 2, complete graphic matroids $(M(K_d), \sigma)$ can be seen to be building blocks for all labeled graph matroid freeness properties. Thus, a first step towards the resolution of this question might be to understand $(M(K_d), \sigma)$-freeness for any pattern $\sigma$, and then study how intersections of such properties behave.

Leaving aside the issue of labeled matroid homomorphisms, another fundamental open problem is whether there must always hold a dichotomy between containment and $\delta$-separation for matroid freeness properties. If this would turn out to be the case, the next question is whether such a dichotomy would extend even further to arbitrary linear-invariant properties.

At the core of all this is the problem of determining when $(M, \sigma)$-freeness and $(N, \tau)$-freeness are identical properties for two syntactically different labeled matroids $(M, \sigma)$ and $(N, \tau)$. If we look at graphic matroids, one observation is that blowing up the underlying graph does not change the property. Formally, for a graph $G$ and for a positive integer $t$, define the *order-$t$ blowup* of $G$ to be the graph $G^{(t)}$ obtained by replacing each vertex of $G$ by an independent set of size $t$ and each edge in $G$ by the complete bipartite graph $K_{t,t}$. Furthermore, if an edge of $G$ is labeled by an element of $\{0,1\}$, then use that label for all edges in the associated complete bipartite graph in the blowup graph; if $\sigma$ was the original labeling for the edges, we call the new labeling for the blowup graph $\sigma^{(t)}$.

The fact that graph blow-ups preserve matroid freeness properties, which we write down as Proposition 6.1 below, is similar in flavor to the Erdős-Stone theorem from extremal graph theory. The Erdős-Stone theorem essentially says that for any graph $G$ and integer $t \geq 1$, $G$-freeeness (i.e., not containing $G$ as an induced subgraph) and $G^{(t)}$-freeness are not $\delta$-separated for any constant $\delta > 0$; see, for example, [Die05]. However, the proof of the analogous statement in the matroid freeness case turns out to be much simpler, because a matroid homomorphism is not required to be injective whereas the subgraph relationship in graphs is an injection.

**Proposition 6.1.** *Given a graph $G$ on $m$ edges and a string $\sigma \in \{0,1\}^m$, suppose $H$ is a subgraph of $G^{(t)}$ for some $t \geq 1$ that contains at least one copy of $G$. Also, suppose $\tau$ is the restriction of $\sigma^{(t)}$ to the edges of $H$. Then, $(M(G), \sigma)$-freeness is identical to $(M(H), \tau)$-freeness.*

The proof is straightforward. The fact that $(M(G), \sigma)$-freeness is contained in $(M(H), \tau)$-freeness follows from Corollary 3.2. The other direction holds because the map that takes each edge of $H$ to the edge of $G$ from where it originated is a labeled matroid homomorphism from $(M(H), \tau)$ to $(M(G), \sigma)$, and so we can apply Lemma 3.1.

It should be noted that Proposition 6.1 is *not* a characterization of equality even for monotone graphic matroid freeness properties. For instance, while $K_4$ is easily seen not to be a subgraph of any blowup of $K_3$, it nevertheless holds that $(M(K_3), 1^*)$-freeness and $(M(K_4), 1^*)$-freeness are identical properties as shown in Proposition 3.3. What our dichotomy theorems in Section 3.3 establish is that for all monotone properties and a nontrivial subclass of non-monotone properties, equality of properties corresponds exactly to existence of matroid homomorphisms in both directions. As noted above, the question whether such a correspondence holds in general for any non-monotone matroid freeness properties remains wide open.

## Acknowledgments

We are grateful to Madhu Sudan for suggesting that we investigate the problems studied in this work, for giving many helpful insights and comments along the way, and for challenging us to redouble our efforts in order to strengthen our first, preliminary results. We also want to thank Asaf Shapira for his collaboration in the early stages of this research, as well as for his continuing useful advice.

## A   Matroids, Matroid Freeness and Systems of Linear Equations

Let us start this appendix by giving a formal definition of what a matroid is for completeness. There are many equivalent ways to define a matroid, and for some of the different formulations it is in fact nontrivial to show that they are equivalent. We will use the definition presented next, and refer the reader to, for instance, [Oxl03, Wil73] for more background on matroid theory.

**Definition A.1 (Matroid).** A *matroid* $M$ is a finite set $S$, along with a set $\mathcal{I}$ of subsets of $S$, such that:

1. The empty set is in $\mathcal{I}$.

2. If $X$ is in $\mathcal{I}$, then every subset of $X$ is also in $\mathcal{I}$.

3. If $X$ and $Y$ are both in $\mathcal{I}$ and $|X| = |Y|+1$, then there exists an element $x \in X \setminus Y$ such that $Y \cup \{x\}$ is in $\mathcal{I}$.

The set $S$ is called the *ground set* of $M$, and the set $\mathcal{I}$ is the collection of *independent sets* of $M$. Those subsets of $S$ which are not in $\mathcal{I}$ are called *dependent*. A maximal independent set — that is, an independent set $X$ which becomes dependent on adding any element of $S$ — is called a *basis* for the matroid. It is a basic result of matroid theory that any two bases of a matroid $M$ must have the same number of elements. This number is called the *rank* of $M$.

Two important classes of matroids, which we briefly discuss next, are *linear matroids* and *graphic matroids*.

We say that a matroid $M$ on a ground set $S = \{x_1, \ldots, x_k\}$ is a *linear matroid*, or *vector matroid*, if there is a field $\mathbb{F}$ and vectors $\mathbf{v}_1, \ldots, \mathbf{v}_k$ in $\mathbb{F}^k$ such that any subset $\{x_i \mid i \in T\}$ indexed by $T \subseteq [k]$ is independent if and only if the corresponding vectors $\{\mathbf{v}_i \mid i \in T\}$ form a linearly independent set. A matroid is *binary* if it is linear with $\mathbb{F} = \mathbb{F}_2$. Note that when we are interested in property testing of linear-invariant properties, the only matroid freeness properties that really make sense to consider are those of linear matroids.

Given a graph $G$, we can let $S$ be the set of edges $E(G)$ of $G$ and $\mathcal{I}$ consist of the subsets of $S = E(G)$ that do not contain any cycles in the graph. Then $M = (S, \mathcal{I})$ can be shown to be a matroid, which we refer to as the *graphic matroid* $M(G)$ over $G$. Any graphic matroid $M(G)$ can be represented as a binary matroid. One way of seeing this is to consider the incidence matrix of $G$ and let $\mathbf{v}_i$ be the rows corresponding to the edges. Then any cycle in $G$ will correspond to a (subset of) vectors summing to zero. Another possibility is to fix any spanning tree of $G$ and let the edges $e_1, e_2, \ldots$ in this spanning tree $T$ correspond to unit vectors $\mathbf{e}_1, \mathbf{e}_2, \ldots$. Then any edge $e$ not in the spanning tree $T$ will correspond to the sum of the vectors for the unique minimal set of edges in $T$ that together with $e$ yields a cycle.

As the reader can see, in this paper we used the latter approach with spanning trees emanating from a single, unique vertex to get our standard representation for $M(K_d)$ (Definition 2.6). Another possibility would have been to use the incidence matrix representation. Note that this would have given a very nice and symmetric representation with all vectors in $M(K_d)$ having Hamming weight 2, and with a basis corresponding to fixing some coordinate $j$ and requiring that all the basis vectors have a 1 in this coordinate. Of course, our standard representation is just taking such a basis and "puncturing" it by deleting the $j^{\text{th}}$ coordinate from all vectors. For our purposes, it somehow turned out that it was very convenient to have a

# A  Matroids, Matroid Freeness and Systems of Linear Equations

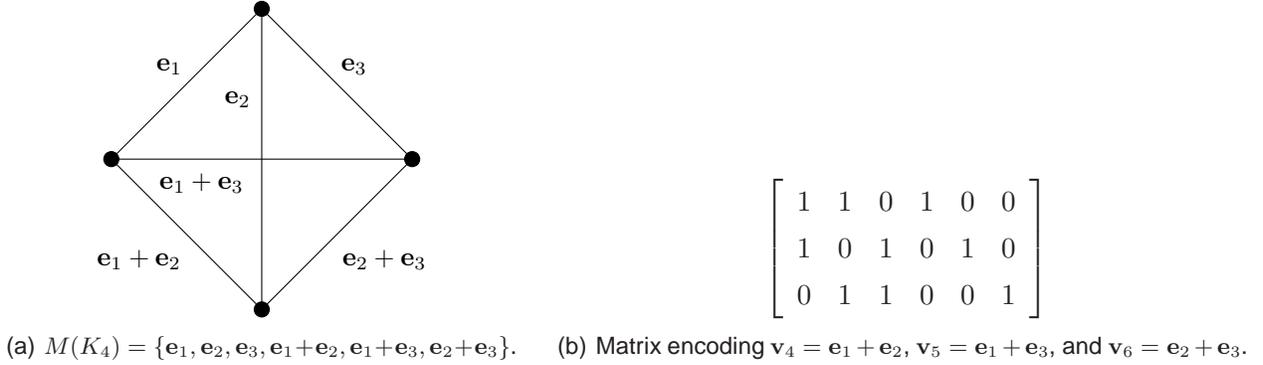

(a) $M(K_4) = \{\mathbf{e}_1, \mathbf{e}_2, \mathbf{e}_3, \mathbf{e}_1+\mathbf{e}_2, \mathbf{e}_1+\mathbf{e}_3, \mathbf{e}_2+\mathbf{e}_3\}$.    (b) Matrix encoding $\mathbf{v}_4 = \mathbf{e}_1 + \mathbf{e}_2$, $\mathbf{v}_5 = \mathbf{e}_1 + \mathbf{e}_3$, and $\mathbf{v}_6 = \mathbf{e}_2 + \mathbf{e}_3$.

**Figure 3:** The binary graphic matroid $M(K_4)$ and the corresponding linear equation system matrix.

representation where all basis vectors and weight 1 and all other vectors had weight 2. However, we just want to point out that this is not the only way of thinking about $M(K_d)$, and that it might be interesting when trying to generalize our results to investigate whether the representation with all vectors of uniform weight 2 might be a more fruitful way of looking at $M(K_d)$.

Let us now explain how one can see that the matroid-freeness representation of properties employed in [BCSX09] and the current paper on the one hand, and the system of linear equations representation of properties used in [KSV09, Sha09, BGS10] on the other, are essentially equivalent. This equivalence is in some sense folklore knowledge, but since it is not entirely obvious a priori, and since we have not seen it actually written down anywhere, we give an explicit exposition of the correspondence here for completeness. Note, however, that we will only discuss monotone properties below. Non-monotone properties have also been formulated using systems of linear equations, in this context most notably in [BGS10], but since the notation is a bit heavier and the ideas are essentially the same, we ignore the issue of non-monotonicity in this appendix for the sake of simplicity.

The *system of linear equations representation* is the following. Let $\mathbb{K}$ be a field. Let $k$ and $\ell$ be fixed integers with $k < \ell$. Let $A\mathbf{x} = \mathbf{b}$ be a system of $k$ linear equations in $\ell$ variables, where $A \in \mathbb{K}^{k \times \ell}$ and $\mathbf{b} \in \mathbb{K}^k$. We say a set $S \subset \mathbb{K}$ is $(A, \mathbf{b})$-*free* if it contains no solution to $A\mathbf{x} = \mathbf{b}$; that is, $S$ is $(A, \mathbf{b})$-free if there is no vector $\mathbf{x} \in S^\ell$ that satisfies all of the $k$ equations in $A\mathbf{x} = \mathbf{b}$.

When considered as a property testing problem, we usually first pick a finite field $\mathbb{F}$ and then take $\mathbb{K} = \mathbb{F}^n$. For properties that arise naturally in mathematics and computer science, it is usually the case that the property can be specified uniformly for all $n$ using a finite description. Thus, it is of particular interest to consider the case when $A$ and $\mathbf{b}$ have finite descriptions. Thus, $A$ is usually taken as having entries over $\mathbb{F}$, not $\mathbb{K}$. Furthermore, in order for the properties to be linear-invariant we take $\mathbf{b} = \mathbf{0}$. (We note, however, that the results in [KSV09, Sha09] also hold when we can pick $A$ (non-uniformly) as any matrix in $\mathbb{K}^{k \times \ell}$ and for any $\mathbb{K}$, not just $\mathbb{K} = \mathbb{F}^n$, and for $\mathbf{b} \neq \mathbf{0}$.)

As a first simple example, the system of linear equations representation corresponding to $(M(C_\ell), 1^*)$-freeness, where $C_\ell$ is the cycle of length $\ell$, consists of just a single linear equation $\sum_{i=1}^\ell x_i = 0$. Therefore we have $A = [1\,1\,1\,\cdots\,1]$ and $\mathbf{b} = \mathbf{0}$, encoding that the sum of $\ell$ vectors is zero. Another simple, but less trivial, example is that of $(K_4, 1^*)$-freeness. Figure 3 shows the graphic matroid $M(K_4)$ and the matrix $A$ its corresponding representation as a system of linear equations.

Let us now consider $(M, 1^*)$-freeness for a general linear matroid $M = \{\mathbf{v}_1, \ldots, \mathbf{v}_{\ell-k}, \mathbf{v}_{\ell-k+1}, \ldots, \mathbf{v}_\ell\}$, where $\{\mathbf{v}_1, \ldots, \mathbf{v}_{\ell-k}\}$ form a basis for the matroid and each of the vectors in $\{\mathbf{v}_{\ell-k+1}, \ldots, \mathbf{v}_\ell\}$ can be written as a linear combination of the first $\ell - k$ vectors $\mathbf{v}_{\ell-k+i} = -\sum_{j=1}^{\ell-k} B_{ij}\mathbf{v}_j$ with coefficients $B_{i,j} \in \mathbb{F}$. Without loss of generality, we can think of the first $\ell - k$ vectors as being $\mathbf{v}_1 = \mathbf{e}_1, \ldots \mathbf{v}_{\ell-k} = \mathbf{e}_{\ell-k}$. To transform $(M, 1^*)$-freeness into a system of linear equations representation, we construct a matrix $A \in \mathbb{F}^{k \times \ell}$ in which the $i^{\text{th}}$ row consists of the coefficients of the linear equations describing $\mathbf{v}_{\ell-k+i}$. Specifically, for every $1 \leq i \leq k$ and $1 \leq j \leq \ell$, $A_{ij} = B_{ij}$ when $1 \leq j \leq \ell - k$; $A_{ij} = 1$ if $j = \ell - k + i$ and $A_{ij} = 0$ otherwise.

To go in the other direction and transform a system of linear equations into matroid freeness representation, we may assume without loss of generality that the matrix $A$ has rank $k$ (otherwise we may delete the redundant rows). Then by permuting the columns of $A$ and appropriately changing the basis of $\mathbb{K}$, we can transform $A$ into the form $[B|I_k]$, where $I_k$ is a $k$-by-$k$ identity matrix and $B$ is a $k$-by-$(\ell - k)$ matrix. Now $B = \{B_{ij}\}$ is exactly the matrix we defined above which contains the coefficients of $k$ linear combinations of the non-basis vectors in the matroid in terms of the $\ell - k$ basis vectors, so it is immediate to recover the matroid freeness representation from this matrix.